\documentclass[
acmsmall]{acmart}

\setcopyright{rightsretained}
\acmJournal{TOSEM}
\acmYear{2024} \acmVolume{1} \acmNumber{1} \acmArticle{1} \acmMonth{1} \acmPrice{}\acmDOI{10.1145/3640332}

\usepackage[utf8x]{inputenc}
\usepackage{booktabs}   

\usepackage{xr}
\usepackage{hyperref}
\usepackage{graphicx}
\usepackage{xcolor}
\usepackage{layouts}

\usepackage{amsmath}

\usepackage[ruled,vlined,linesnumbered]{algorithm2e}

\usepackage{mathtools}
\usepackage{dsfont}
\usepackage{paralist}
\usepackage[noabbrev,capitalize]{cleveref}

\usepackage{textcomp}

\usepackage{pifont}
\usepackage{tikz}
\usepackage{pgfplots}
\usetikzlibrary{calc}

\usepackage{float}

\usepackage{makecell}
\usepackage{multirow}

\usepackage{siunitx}
\DeclareSIUnit{\pp}{\textup{pp}}

\usepackage[font=small,labelfont=bf]{caption}
\usepackage{subcaption} 

\begin{document}
\title[Rigorous Assessment of Model Inference Accuracy]{Rigorous 
Assessment of Model Inference Accuracy using Language Cardinality}

\author{Donato Clun}
\orcid{0000-0001-5190-8957}
\affiliation{\institution{Imperial College London}
  \country{UK}
}
\email{d.clun@imperial.ac.uk}
\authornote{Part of this work was done while the author was a visiting PhD Student at the University of Luxembourg, Luxembourg.}
\author{Donghwan Shin}
\orcid{0000-0002-0840-6449}
\affiliation{\institution{University of Sheffield}
  \country{UK}
}
\email{d.shin@sheffield.ac.uk}
\authornote{Part of this work was done while the author was affiliated
  with the University of Luxembourg, Luxembourg.}
\author{Antonio Filieri}
\orcid{0000-0001-9646-646X}
\affiliation{\institution{Imperial College London}
  \country{UK}
}
\email{a.filieri@imperial.ac.uk}

\author{Domenico Bianculli}
\orcid{0000-0002-4854-685X}
\affiliation{\institution{University of Luxembourg}
  \country{Luxembourg} 
}
\email{domenico.bianculli@uni.lu}

\begin{abstract}
  Models such as finite state automata are widely used to abstract the behavior
  of software systems by capturing the sequences of events observable during 
  their execution. Nevertheless, models rarely exist in practice and, when 
  they do, get easily outdated; moreover, manually building and maintaining 
  models is costly and error-prone. As a result, a variety of model inference
  methods that automatically construct models from execution traces have been
  proposed to address these issues.

  However, performing a systematic and reliable accuracy assessment of inferred models
  remains an open problem. Even when a reference model is given, most existing
  model accuracy assessment methods may return misleading and biased results. This 
  is mainly due to their reliance on statistical estimators over a finite 
  number of randomly generated traces, introducing avoidable uncertainty 
  about the estimation and being sensitive to the parameters of the random 
  trace generative process.
  
  This paper addresses this problem by developing a systematic approach based
  on analytic combinatorics that minimizes bias and uncertainty in model 
  accuracy assessment by replacing statistical estimation with deterministic
  accuracy measures. We experimentally demonstrate the consistency and 
  applicability of our approach by assessing the accuracy of models inferred 
  by state-of-the-art inference tools against reference models from 
  established specification mining benchmarks.
\end{abstract}



\maketitle

\section{Introduction} 
Software system models, typically in the form of Finite State Automata
(FSA), have been widely used to abstract the behavior of software
components and their interactions. Such models are important in many
applications, including test data generation~\cite{6200086}, model
checking~\cite{clarke2018model}, and program
comprehension~\cite{Cook:1998:287001}. Nevertheless, they are rarely
available during software development and, if they do, get easily
outdated. This is mainly because manually building and maintaining
such models is both time-consuming and error-prone. To address this
problem, a variety of \emph{model inference} algorithms have been
proposed~\cite{biermann1972synthesis,Beschastnikh2011Lev,Walkinshaw2016,Emam2018Inf,mariani2017gk}. 
These algorithms automatically extract a system's behavior model by capturing visible 
events generated during the system's execution.

\emph{Model assessment}, i.e., assessing the accuracy of inferred models, 
is an essential task to evaluate and compare different model inference 
algorithms. When reference models are given as ground truth, model 
assessment seems straightforward; for example, one can compute the similarity 
between the languages defined by the inferred and reference models. However, 
the languages are often infinite, making it complex to measure the degree 
of their similarity.

To address this issue, it is common to rely on statistical accuracy
estimation. For example, the idea behind a popular method, called
\emph{trace similarity}~\cite{Lo2006}, is to generate sampled 
traces\footnote{A \emph{trace} is the sequence of events generated during the 
system's execution. Also known as \emph{run}, 
\emph{sequence}, \emph{execution trace.}}
by randomly traversing both reference and inferred models, and estimate the 
accuracy by checking how many of the traces generated using one model are 
accepted by the other. The idea is that a high number of accepted sampled 
traces is an indicator of a high degree of 
similarity. 
Although this idea is very intuitive, the random traverse can introduce a large
degree of uncertainty about the estimation, depending on the parameters of the
random trace generative process (e.g., the probability of ending the traverse in
an accepting state rather than continuing through one of its outgoing
transitions), and the topology of the model (e.g., a model may have features
that are hard to exercise through random exploration). While there is no
intrinsic probability distribution over the traces accepted by a finite state
model, the random trace generation does induce such a distribution, therefore
introducing an evaluation bias: the measured accuracy will reflect how well the
inferred model can classify traces drawn according to that distribution.

In an attempt to reduce the uncertainty due to the random traversal, an
alternative consists in using deterministic trace generation
methods~\cite{Walkinshaw2008}. These generate sets of traces that are guaranteed
to cover all the models' behaviors without favoring any specific one.
Although this increases the reproducibility and the chances of revealing
discrepancies, there are still remaining issues. First, each discrepancy (a
trace upon which the reference and the inferred model disagree) may be representative of
a larger or smaller class of traces, thus making it difficult to quantify its
impact. Second, these methods often generate sets of traces containing a
disproportionately large amount of traces not accepted by the reference model,
which may lead to skewed accuracy results. Third, the generated set of
traces can become exceedingly large if the difference between the number of
states in the reference and the inferred model is large, which is a common occurrence in
model inference, hindering its practical usability. 

In this paper we propose a method to rigorously measure the
accuracy of an inferred model against a reference one (both in the form of
FSA) by considering all the possible traces up to an arbitrary
finite maximum trace length.
The maximum trace length considered is a parameter set by the user, which can be
set to a value relevant for the application under analysis, to a value
large enough to guarantee that all the behaviors of the reference and inferred
models are exercised, or to an arbitrarily large value, effectively computing
the asymptotic accuracy. 

Our method is deterministic (i.e., for a given pair of reference and
inferred models it always returns the same result), does not depend on the model
structure (i.e., assessing against the same reference model different models
accepting the same language will generate the same accuracy measurement), and
does not introduce any evaluation bias other than the maximum trace length
(i.e., the computed metrics accurately describe how well the inferred model can
classify traces drawn with uniform distribution from the set of all the traces
up to the maximum length).

Within this paper we also highlight how an inferred model can show a variable
level of accuracy depending on the length of the traces used in the assessment.
This is an aspect that is generally not considered by current assessment
methodologies. Following this observation, we propose an additional assessment
method that computes a pair of precision and recall values for each
trace length, within a range specified by the user. Each value considers all 
the possible traces of a given length.

Central to our solution is the use of analytic combinatorics to count the number
of traces in a (possibly infinite) regular language, up to an arbitrary maximum
trace length, without explicitly enumerating them. In this regard, to increase
the practical applicability of our approach, we improved the scalability of the
analytic combinatorics approaches currently used to compute the cardinality of
regular languages~\cite{Flajolet2009}. 

We have implemented the proposed method in a prototype and experimentally evaluated
it using reference models previously used in the model inference literature, and
inferred models generated using well known model inference methods.
The experimental results obtained when assessing the applicability of
our method on the real-world model show that it is scalable enough to be used in practice.
We have also compared the generated accuracy measurements with the results obtained
using other popular assessment methods. The results indicate differences
caused by the evaluation bias introduced by current assessment methods, and
show that our proposed approach can be used to address this issue.

\paragraph{Significance}
Assessing the accuracy of inferred models against ground truth reference models
is crucial in evaluating newly developed model inference methods, as it serves
as the primary method for evaluating effectiveness and comparing different
inference techniques. Our work is targeting researchers involved in the
development or evaluation of model inference methods. Our goal is to provide a tool to evaluate \emph{inference
algorithms}, rather than the inferred models. In particular, we highlight how
currently commonly used assessment methods can generate misleading results, thus
interfering with the development of effective inference techniques and creating
false expectations about the quality of the inferred models. Furthermore, through
our novel model assessment method, we provide more accurate and comprehensive
insights about the model assessment results, including new insights
describing how the model accuracy differs when evaluating it on traces of
different lengths.

\paragraph{Contributions} In summary, this paper makes the following contributions:
\begin{itemize}
    \item A novel model assessment method that measures the precision and
      recall values of an
        inferred model, with respect to a reference ground truth model,
        considering all the traces up to a finite maximum trace length. This method is:
        \begin{itemize}
            \item Deterministic: repeated executions return the same result.
            \item Comprehensive: it considers all the traces up to the maximum length.
            \item Unbiased: all the traces considered in the evaluation have the same 
            weight on the result.
            \item Model-independent: assessing different models
            accepting the same language against the same reference model generates the same result. 
        \end{itemize}
    \item A further development of the assessment method, measuring the model accuracy
        separately for each trace length, over a given range.
    \item An empirical comparison of the accuracy measurement obtained using our
        methods with measurements generated using other popular assessment
        methods.
    \item An experimental evaluation of the applicability of our method.
    \item An improvement of currently used analytic combinatorics approaches to
        compute the cardinality of regular languages, up to a finite maximum
        trace length.
\end{itemize}

\paragraph{Outline}
The rest of the paper is organized as follows. 
In Section~\ref{sec:background}, we present a brief background on the theory of
formal languages, model assessment, and analytic combinatorics as a tool to
compute the cardinality of regular languages. 
In Section~\ref{sec:existing-approaches}, we discuss two prominent classes of
model assessment methods from the literature: statistical estimation
(the most commonly used accuracy assessment method) and model-based (developed
to address shortcomings of the first), reflecting on the open issues that
motivated this work.
Section~\ref{sec:our_approach} presents our main contribution: 
a novel assessment method,
based on analytical measures for the cardinality of languages,
that addresses the issues highlighted in the preceding section,
and a further development that allows to evaluate how the model
accuracy changes over a range of trace lengths.
To increase the practical applicability of our method, in
Section~\ref{sec:fastOGF}, we present an improvement of currently used analytic
combinatorics approaches to compute the cardinality of regular languages.
In Section~\ref{sec:experimentalEval} we evaluate our method experimentally,
focusing on whether it is suitable to replace the assessment methods discussed
in Section~\ref{sec:existing-approaches}, and how the different methods'
assessment results compare.
Section~\ref{sec:related-work} presents relevant related work, 
and Section~\ref{sec:conclusions} concludes the paper with future work directions.

\section{Background}\label{sec:background}

\subsection{Models and Languages}\label{sec:background_models_and_languages}
In this paper, we consider models in the form of Deterministic Finite-state 
Automata (DFA). A DFA is a tuple $\mathcal{A} = (\Sigma, Q, q_0, F, \delta)$,
where $\Sigma$ is a finite alphabet, $Q$ is the set of states, $q_0 \in Q$ is the 
initial state, $F \subseteq Q$ is the set of accepting states, and 
$\delta : Q \times \Sigma \rightarrow Q$ is the transition function.
A \textit{trace} is a finite sequence $t = \langle \sigma_1 \sigma_2 \dots \sigma_n \rangle$ 
of elements $\sigma_i \in \Sigma$ for $i = 1, \dots, n$.  
A trace $t = \langle \sigma_1 \sigma_2 \dots \sigma_n \rangle$ is \emph{accepted} by 
$\mathcal{A}$ if there exists a sequence of states $\langle q_0, q_1, \dots, q_n \rangle$ such
that (1) $q_i \in Q$ for $i=1, \dots, n$, (2) $\delta(q_{i-1}, \sigma_i) = q_i$
for $i=1, \dots, n$, (3) $q_0$ is the initial state, and (4) $q_n \in F$.
A state $q \in Q$ is an \emph{error state} if no accepting state is reachable 
from $q$. 
Let $\Sigma^*$ be the set of all possible traces over $\Sigma$ (including the empty trace).
The \textit{language accepted} by $\mathcal{A}$, denoted by 
$\mathfrak{L}(\mathcal{A}) \subseteq \Sigma^*$, 
is the set of all traces accepted by $\mathcal{A}$.
Two DFAs are \textit{equivalent} if they accept the same language.

DFAs accept regular languages, which are closed under union, intersection, 
and complement. These operations on regular languages correspond to analogous
operations on the automata accepting the languages.
Therefore, with abuse of notation, we will use the union ($\cdot \cup \cdot $), 
intersection ($\cdot \cap \cdot$), and complement ($\overline{\cdot}$) 
operators on both automata and the languages they accept; for example, 
$\mathcal{A} \cup \overline{\mathcal{B}}$ represents the DFA accepting 
any trace that is accepted by $\mathcal{A}$ or not accepted by $\mathcal{B}$.

\subsection{Model Assessment as Language Comparison}\label{sec:model-evaluation}

Given a reference model $\mathcal{R}$ and an inferred model $\mathcal{H}$ 
over the same alphabet $\Sigma$, to assess the accuracy of the inferred model 
we need a measure of how well the language accepted by $\mathcal{H}$ approximates the
language accepted by $\mathcal{R}$. Drawing from established theory on 
classification assessment methods~\cite{tharwat2020classification}, 
$\Sigma^*$ can be partitioned  into four subsets of traces: 
True Positives ($\Sigma^*_{\mathit{TP}} = \mathfrak{L}(\mathcal{R} \cap \mathcal{H})$), 
True Negatives ($\Sigma^*_{\mathit{TN}} = \mathfrak{L}(\overline{\mathcal{R}} \cap \overline{\mathcal{H}})$), 
False Positives ($\Sigma^*_{\mathit{FP}} = \mathfrak{L}(\overline{\mathcal{R}} \cap \mathcal{H})$), and 
False Negatives ($\Sigma^*_{\mathit{FN}} = \mathfrak{L}(\mathcal{R} \cap \overline{\mathcal{H}})$). 

By generalization, any set of traces $E \subseteq \Sigma^*$ (which we call \textit{evaluation set}) 
can be partitioned into four subsets $E_{\mathit{TP}}$, $E_{\mathit{TN}}$, $E_{\mathit{FP}}$, 
and $E_{\mathit{FN}}$, containing true positives, true negatives, false positives and false negatives, respectively. 
If the evaluation set is finite, several accuracy metrics can be defined 
based on the \emph{relative cardinality} ($|\cdot|$) of these four languages. 
In this paper, we will focus on precision and recall:
\begin{equation}\label{eq:precision-recall-background}
    \mathit{precision} = \frac{|E_{\mathit{TP}}|}{|E_{\mathit{TP}}| + |E_{\mathit{FP}}|} ; \quad 
    \textit{recall} = \frac{|E_{\mathit{TP}}|}{|E_{\mathit{TP}}| + |E_{\mathit{FN}}|}
\end{equation}
which are widely used in model inference literature~\cite{Krka2014,Lo2012,Lo2006}.

It is worth noting that precision and recall cannot be used as distance functions 
in a metric space as they are not symmetric by definition.
However, if needed, the model assessment approach discussed in this work can be adapted to
compute the Jaccard distance (i.e., distance between $\mathcal{R}$ and $\mathcal{H}$)
with respect to a finite evaluation set (\emph{bounded Jaccard}), yielding a metric space.

In the rest of this paper we will continue to use $\mathcal{R}$ and $\mathcal{H}$ 
to denote the reference and the inferred model respectively, both assumed to be defined 
over the same alphabet $\Sigma$.

\subsection{Analytic Combinatorics and Cardinality of Regular
Languages}\label{sec:cardinality_analytic_combinatorics} 

Analytic combinatorics~\cite{Flajolet2009} is a theory used to build, manipulate and 
analyze exact enumerative descriptions of combinatorial structures, typically
focusing on structures whose realizations are too many for explicit enumeration.
\emph{Generating functions} are a core tool in analytic combinatorics, as they
enable to concisely define and operate on certain infinite sequences.
Let us consider a discrete, possibly infinite sequence of real values $a_n$, with 
$n=0, 1, 2, \cdots$. An \emph{ordinary generating function} (OGF) is a mathematical 
object encoding the sequence as the coefficients of a formal power series: 
\begin{equation}
	f(z) = a_0 + a_1 z + a_2 z^2 + a_3 z^3 + \cdots = \sum_{n=0}^{\infty} a_n z^n	
\end{equation}
Given an OGF $f(z)$, the $n$-th coefficient $a_n$ can be retrieved as the $n$-th 
Taylor coefficient of $f(z)$
\begin{equation}
	a_n = \frac{1}{n!}\frac{\partial^n}{\partial z^n} f(z) \big|_{z=0}	
\end{equation}

A regular language $L$, accepted by a DFA $\mathcal{A}$, is a combinatorial 
structure representing all the traces belonging to the language. 
The \emph{sequence of cardinalities} of $L$ is the sequence of number of traces 
$a_n$ of length $n=0,1,2,\cdots$ accepted by $\mathcal{A}$, i.e., the 
cardinality of the language accepted by $\mathcal{A}$ restricted to traces of 
length $n$, $\mathfrak{L}(\mathcal{A}) \cap \Sigma^n$.
For our purpose, we aim at constructing an OGF denoted with $\mathfrak{G}_L(z)$ encoding 
in a compact way the sequence of cardinalities of $L$.

The OGF for the sequence of cardinalities of a regular language is always a 
rational function~\cite{Flajolet2009}.
For example, let us consider the language $L=\Sigma^*$, with $\Sigma=\{0, 1\}$. 
The empty trace, with length 0, is accepted ($a_0=1$); two traces of length 1 
are accepted ($a_1=2$); four of length 2 ($a_2=4$); and so on. The infinite 
sequence $a_n=\{1, 2, 4, 8, 16, \cdots\}$ can be compactly encoded by the 
OGF $\mathfrak{G}_L(z)=\frac{1}{1-2z}$. 

Given a regular language $L$, computing the OGF $\mathfrak{G}_L(z)$ 
is typically reduced to the problem of counting the number of accepting paths 
of length $n$ of the minimal automaton accepting $L$, via the established 
\emph{transfer matrix method}~\cite{Flajolet2009,stanley2011enumerativeCombinatorics}. 
This algorithm relies on linear algebraic operations on the matrix representing 
the transition relation of the automaton~\cite[Proposition I.3]{Flajolet2009}. 
While the transfer matrix method works in general, in Section~\ref{sec:fastOGF} we 
will present an alternative, equivalent algorithm to obtain the
OGF of a regular language that significantly outperformed the transfer matrix 
method in our tests.

Once the OGF $\mathfrak{G}_L(z)$ has been computed using one of the above methods, the values of the sequence 
$a_i, i\geq 0$ can be recovered as the Taylor coefficients of 
$\mathfrak{G}_L(z)$. 
Fortunately, since the OGF for the sequence of cardinalities of a 
regular language is always a rational function~\cite{Flajolet2009} which 
can be written as the quotient between two finite degree polynomials
$
	\mathfrak{G}_L(z) = \frac{N(z)}{D(z)} 
$,
it is possible (and faster) to derive a recurrence relation that allows
computing the values of the sequence without performing symbolic
differentiation. Let $N(z)$ and $D(z)$ be polynomials of maximum degree $m$,
and recall that the OGF is equal to the formal power series with coefficients
$a_i$.
\begin{equation}
	\frac{N(z)}{D(z)} = 
	\frac{b_0 + b_1 z + \cdots + b_m z^m}{c_0 + c_1 z + \cdots + c_m z^m} =
	\sum_{i=0}^{\infty} a_i z^i
\end{equation}

By multiplying by $D(z)$ and expanding the sum, one obtains
\begin{equation}
	b_0 + b_1 z + \cdots + b_m z^m = (c_0 + c_1 z + \cdots + c_m z^m)(a_0 + a_1 z + a_2 z^2 + \cdots)
\end{equation}
and by equating the coefficients, 
(considering $b_n$ and $c_n$ equal to zero if $n > m$),
one obtains 
$
	b_n = \sum_{j=0}^{n} c_j a_{n-j} 
$
and therefore
$
	a_n = \frac{b_n - \sum_{j = 1}^{n} c_j a_{n-j}}{c_0}
$
which defines the sequence $a_n$.

 \section{Existing methods for Model Assessment}\label{sec:existing-approaches}

Following Equation~\eqref{eq:precision-recall-background}, when a finite evaluation 
set of traces $E$ is available, measuring the accuracy of 
$\mathcal{H}$ against $\mathcal{R}$ using precision and recall can be 
achieved by comparing the acceptance of each trace in $E$. However, obtaining a finite 
and representative evaluation set over $\Sigma^*$ is a challenging problem. 
In this section, we discuss the two prominent classes of methods from
the literature 
to generate evaluation sets for model assessment.
Methods in the first class --- \emph{Statistical Estimation} --- generate the evaluation 
set of traces by means of a random walk over the automata's paths. Methods in 
the second class --- \emph{Model-Based} --- use a deterministic traversal of 
the automata's transition relation to generate an evaluation set that is
likely to reveal divergences between the reference model and the
inferred one.

\subsection{Statistical Estimation using Random Walks}\label{sec:evaluation_with_trace_similarity}
Statistical estimation~\cite{Busany2019,Lo2012,Lo2009,Walkinshaw2013,Walkinshaw2013stamina,Mariani2010}
represents a class of model assessment methods based on finite evaluation multisets (denoted by $E$) 
of traces randomly generated (sampled) over $\Sigma^*$.
The random process used to generate $E$ is the characterizing element of each method. 
The most popular statistical estimation method is \emph{trace similarity}, proposed by 
\citet{Lo2006}, which generates $E$ by performing \emph{random walks} on 
the reference and inferred models, rather than generating random sequences from $\Sigma^*$. 

The core random walk algorithm is defined as follows, starting from the initial
state. If the current state is accepting, the procedure randomly decides whether
to terminate the walk or to continue; if the walk is not terminated, it randomly
selects one among the outgoing transitions from the current state, adds the
transition symbol to the trace, updates the current state to the target of the
transition, and then the random walk continues from the new current state. If an
error state\footnote{As defined in
Section~\ref{sec:background_models_and_languages}, an error state is a state from which no
accepting state can be reached.} is reached, the current trace is discarded and
the random walk restarts from the initial state. 

To compute the \emph{precision} using trace similarity, an evaluation multiset $E$
is generated by performing random walks on the inferred model $\mathcal{H}$, therefore
generating traces that are either true positives or false positives. Precision
is then defined as the ratio of traces in $E$ accepted also by the reference
model $\mathcal{R}$, i.e.\ the proportion of true positives\footnote{The
rationale for this process stems from the observation that $\frac{|TP|}{|TP| +
|FP|} = \frac{|\mathfrak{L}(\mathcal{R} \cap
\mathcal{H})|}{|\mathfrak{L}(\mathcal{R} \cap \mathcal{H})| +
|\mathfrak{L}(\overline{\mathcal{R}} \cap \mathcal{H})|} =
\frac{|\mathfrak{L}(\mathcal{R} \cap
\mathcal{H})|}{|\mathfrak{L}(\mathcal{H})|}$, which on the surface reminds of
the conditional probability of a trace being accepted by $\mathcal{R}$ given
that the trace is accepted by $\mathcal{H}$. We shall demonstrate in the
remaining of this section that this conditional probability interpretation is
itself conditional on the random walk process, which may render the computed
precision meaningless.}.
Similarly, to compute the \emph{recall}, the evaluation multiset $E$ is generated by
performing random walks on the reference model $\mathcal{R}$, generating traces
that are either true positives or false negatives. Recall is then defined as the
proportion of traces in $E$ accepted also by the inferred model $\mathcal{H}$.
In both cases, model coverage can be achieved by repeated, independent random
walks, e.g., until each transition is traversed a minimum number of times.

It is important to note that, when~\citet{Lo2006} originally proposed the trace
similarity method, it was assumed either (a) that the models are probabilistic and
the random walk is performed according to the transition probabilities of the
models, or (b) that the models are deterministic and the user specifies how the
random choices necessary to generate traces are made (defaulting to a uniform
distribution over the outgoing transitions if not otherwise specified). The
transition probability distribution is an important aspect of the assessment
because, as we will soon discuss, it affects the results. In the model
inference literature, the use of trace similarity on deterministic models, i.e.,
whose transition probabilities are not specified by the model itself, is
predominant. This is because the inference methods that are most well-known
(e.g., $k$-tails~\cite{biermann1972synthesis}) or considered state-of-the-art (e.g.,
MINT~\cite{Walkinshaw2016}) generate deterministic models. Moreover, the
reference models used in the assessment are also most often deterministic,
because they are either manually created (for example, from API documentation and
reference books~\cite{Pradel2010}) or they come from a formal
specification of the target language, thus with no information about the
probability distribution of the traces. 

On the other hand, a random walk algorithm does require a probability distribution 
to select which transition to traverse at each step, and to decide whether to continue
or to terminate the random walk when an accepting state is reached.

To decide which transition to follow, it is common practice to select one of the
possible alternatives with uniform probability distribution~\cite{Lo2006}. For
the decision of whether to continue or terminate the random walk once a final
state is reached, a number of strategies are commonly used; e.g.,
\citet{Walkinshaw2013stamina} define a termination probability that is inversely
proportional to the number of outgoing transitions.

A factor that is often overlooked in the statistical estimation of the model
accuracy is how the (arbitrary) randomness of the trace generation impacts the model assessment. 
Inferred and reference models merely represent languages, i.e., 
(possibly infinite) sets of traces over a finite alphabet. 
There is no intrinsic probability distribution over the traces of a finite 
state model. On the other hand, the random trace generation process used to 
produce the evaluation multiset (e.g., a random walk, in the case of trace 
similarity) induces a probability distribution over the accepted language. 
This distribution may be non-uniform (i.e., different traces may be generated with different
probabilities) therefore introducing a \emph{sampling bias} in the
model assessment: the precision 
and recall values obtained using this random sampling will reflect how well the 
inferred model can classify traces \textit{drawn according to that distribution}.
Unless this distribution reflects domain knowledge about the 
application being analyzed (i.e., if the models are purposefully probabilistic, 
and they describe the probability that different 
traces have of being generated by the system under analysis), this sampling 
bias is not desirable.

For example, let us consider a random walk using a fixed \textit{termination probability} 
$p_a$ to decide the trace termination, and uniform sampling among the available 
alternatives to decide which transition to follow.
Intuitively, the termination probability has an impact on the length of the sampled traces:
increasing $p_a$ makes shorter traces more likely to be sampled (correspondingly, 
longer traces less likely). However, 
it is hard to predict the actual length distribution without 
taking into account the specific topology of the model from which the traces are sampled. 
Moreover, the topology of the model also induces a non-uniform distribution among 
accepted traces of the same length: the likelihood of a trace 
depends on the probability of selecting each of its transitions in a given order, with
each of these selections generally depending on the number of successors of the source state.

We have thus identified two sampling biases in the random walk:
\begin{itemize}
	\item shorter traces are more likely to be sampled (but exactly how likely is topology-dependent);
	\item for a given trace length, some traces are more likely to be sampled than others.
\end{itemize}
When a random walk is used to generate the evaluation multiset for trace similarity,
both sampling biases become systematic faults of the method, and their impact 
on the model assessment is difficult to quantify a priori since both
depend on specific random walk parameters and the topology of the model 
to which the random walk is applied. Furthermore, since for trace similarity methods 
precision and recall are measured using random walks on different models 
(i.e., $\mathcal{H}$ for precision, $\mathcal{R}$ for recall), the assessment  
mixes two \emph{different sampling biases}, making the two measures incomparable 
and generally impossible to aggregate.

\begin{figure}[t]
	\centering
	\begin{subfigure}[b]{0.4\textwidth}
		\centering
		\includegraphics[width=\textwidth]{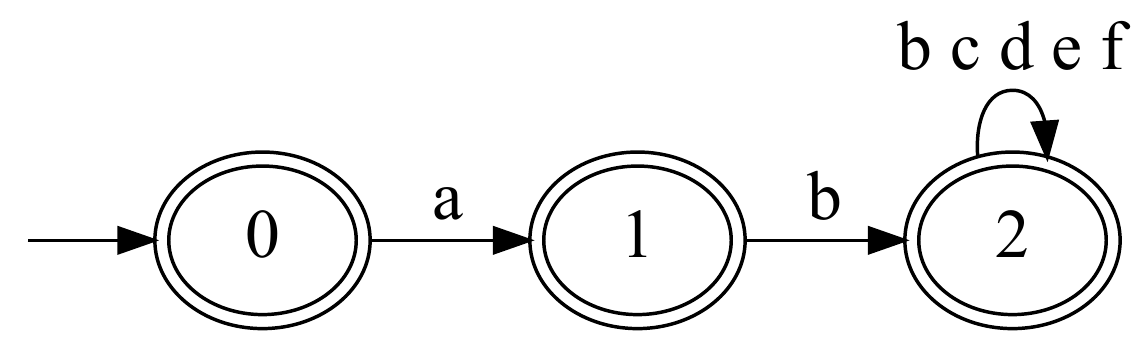}
		\caption{Reference model $\mathcal{R}$.}\label{fig:ref-model-trace-sim}
	\end{subfigure}
	\hfill
	\begin{subfigure}[b]{0.27\textwidth}
		\centering
		\includegraphics[width=\textwidth]{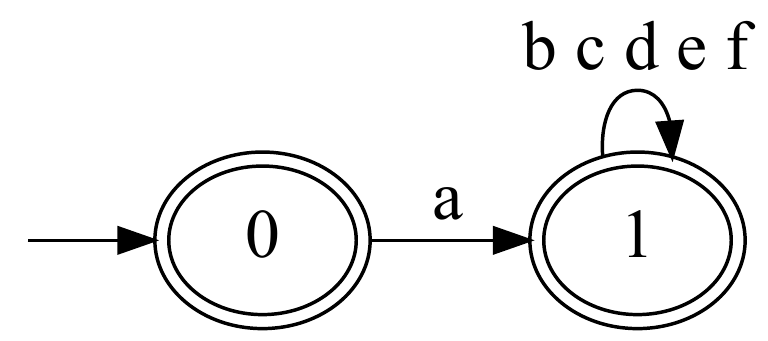}
		\caption{Inferred model $\mathcal{H}$.}\label{fig:inferred-model-trace-sim}
	\end{subfigure}
    \caption{Example of reference and inferred models}\label{fig:trace-sim-failure}
  \end{figure}

\begin{figure}[h]
	\centering
    \includegraphics[]{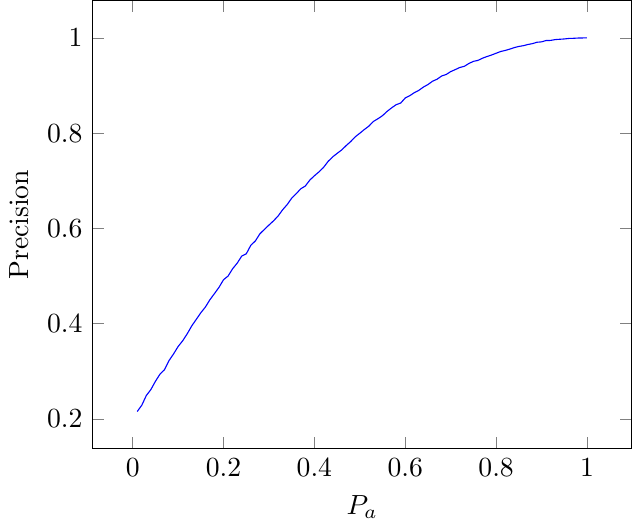}
    \caption{Sensitivity of statistical estimation to changes in the
      random walk for the models in Figure~\ref{fig:trace-sim-failure}}\label{fig:trace-sim-failure-data}
\end{figure}

Let us show with an example how sensitive the model's accuracy calculated by 
trace similarity is to the choice of random walk 
parameters. We take a concrete example based on the reference model 
\texttt{Signature}, adapted from the benchmark models provided in~\citet{Krka2014}. 
For readability, we used single symbols on transition labels 
and omitted transitions to error states. The reference model $\mathcal{R}$ is
shown in Figure~\ref{fig:ref-model-trace-sim}, while Figure~\ref{fig:inferred-model-trace-sim}
shows the inferred model $\mathcal{H}$, obtained with MINT~\cite{Walkinshaw2016} ---
a state-of-the-art model inference tool --- with its default configuration, on 
a set of 100 traces accepted by $\mathcal{R}$. To see the impact of different parameters of random 
walks on the precision of $\mathcal{H}$, we vary the termination probability (denoted with $p_a$) 
between $0.01$ and $1$, while the choice of the outgoing transition is always done uniformly 
among the available alternatives. For each value of $p_a$, 50,000 random traces are 
generated to assess the precision of $\mathcal{H}$. 
Figure~\ref{fig:trace-sim-failure-data} shows the result.
When $p_a=1$, the random walk always generates a trace of length zero (since 
the empty string is accepted by $\mathcal{H}$), which is a true positive, and 
therefore the precision value converges to 1. On the contrary, as $p_a$ 
decreases, longer traces are generated and the precision value converges to 0.2.
In this simple example, this range of values could have been predicted analytically 
because, for traces 
of length $l$ ($l \geq 2$), the number of true 
positives is $5^{(l-2)}$ (the set of true positives is the language 
determined by the regular expression $a \cdot b \cdot \{b, c, d, e, f\}^*$)
while the number of false positives is $4\cdot5^{(l-2)}$ (the set of 
false positives is the language determined by the regular expression 
$a \cdot \{c, d, e, f\} \cdot \{b, c, d, e, f\}^*$), and therefore 
$\lim_{l\to\infty}\frac{5^{(l-2)}}{5^{(l-2)} + 4\cdot5^{(l-2)}} = 0.2$.
As a result, we can see that the precision value varies between 0.2 and 1.0 
depending on the value of $p_a$, implying that the result of trace similarity 
is extremely sensitive to the choice of the random walk parametrization.
Unlike in this simple example, however, the effect of the random walk on the 
assessment result is in general difficult to predict analytically.

\subsubsection{Statistical Estimation by Sampling $\Sigma^*$}\label{sec:sampling-sigma-star}
The sampling bias caused by the random trace generation is fundamentally unavoidable,
because the sampled languages are in general infinite and there exists no uniform 
distribution over a discrete infinite set.
There are, however, alternative methods to generate a random evaluation multiset $E$
(and then compute precision and recall using equation~\eqref{eq:precision-recall-background})
that allow for more control over the sampling bias.

The simplest method consists in generating traces by combining symbols randomly 
selected from the alphabet $\Sigma$, without considering the models under 
analysis, thus making the sampling bias model-independent.
This can be done, for example, by firstly
selecting a random trace length $l$, and then concatenating $l$ symbols, each 
one randomly selected from $\Sigma$. The generated trace is then added to the 
evaluation multiset $E$, and the process is repeated until a stopping condition is 
met (e.g., a certain number of traces has been generated).
Note that, when the maximum length of a sample trace $n$ is fixed, 
it is possible to perform this trace generation in a way that 
ensures uniform probability distribution over $\Sigma^{\leq n}$: 
each trace length $l \leq n$ must be chosen with probability proportional to 
$|\Sigma^l|$, and each symbol must be chosen with uniform distribution.

Despite its simplicity and its ability to reduce the sampling bias,
compared to methods based on random walks like trace similarity,
sampling from $\Sigma^n$ is rarely used in practice due to its computational cost: 
for most practical problems and large values of $n$, the proportion of $\Sigma^n$ 
accepted by either $\mathcal{H}$ or $\mathcal{R}$ tends to decrease with $n$, leading to 
most samples being true negatives, whose number does not contribute to the computation 
of either precision or recall. We will show instances of this problem experimentally in 
Section~\ref{sec:experimentalEval}.

 \subsection{Model-Based Assessment}\label{sec:evaluation_with_model_based_testing}
Walkinshaw et al.~\cite{Walkinshaw2008} proposed a new method based on
methodologies from \textit{model-based testing} (MBT), aimed at mitigating the sampling
bias introduced by the random sampling and increasing the reproducibility of
the results, as it is deterministic and thus not affected by the statistical
uncertainty that comes from using a finite randomly generated evaluation set.
The intuition is that MBT methods can deterministically generate a set 
of traces that is \emph{comprehensive} (in the sense that any erroneous behavior
of the inferred model would be detected) that can be used as evaluation set.

There are multiple MBT methods for DFAs~\cite{Chow1978,Bonifacio2008} that,
given a reference model $\mathcal{R}$ and an upper bound on the number of
states the 
inferred model ($\mathcal{H}$ in our case) is allowed to contain, can check
the equivalence of $\mathcal{R}$ and $\mathcal{H}$ by testing them on a finite 
set of traces,
i.e., it can generate a set of traces $T$ such that, ideally,
$\mathfrak{L}(\mathcal{R}) = \mathfrak{L}(\mathcal{H}) \Leftrightarrow \forall t
\in T : \mathcal{R}(t) = \mathcal{H}(t)$. The idea proposed by Walkinshaw et
al.~\cite{Walkinshaw2008} is to compute precision and recall using the set of
traces $T$ (generated using the MBT method of choice) as evaluation set $E$.

As an example, let us describe how the W-method~\cite{Chow1978}, a widely-used
MBT method for DFA, generates a set of tests $T$. Given the reference model
$\mathcal{R} = (\Sigma, Q_{\mathcal{R}}, q_{i_\mathcal{R}}, F_{\mathcal{R}},
\delta_{\mathcal{R}})$ and an upper bound to the number of states of $\mathcal{H}$
denoted with $m$, it constructs two sets of traces: a \textit{state cover}
(denoted with $C_{\mathcal{R}}$) and a \textit{characterization set} (denoted
with $D_{\mathcal{R}}$). The \textit{state cover} $C_{\mathcal{R}}$ is a
prefix-closed subset of $\Sigma^*$ containing traces reaching each state of
$\mathcal{R}$, i.e., $\forall q \in Q_{\mathcal{R}} \; \exists t \in
C_{\mathcal{R}} : \delta_{\mathcal{R}} (q_{i_\mathcal{R}}, t) = q$. 
The \textit{characterization set}
$D_{\mathcal{R}}$ is a subset of $\Sigma^*$ such that, for any pair of states
$q_a, q_b \in Q_{\mathcal{R}}$ with $q_a \neq q_b$, it contains a
\textit{distinguishing trace} $t$ such that $\delta_{\mathcal{R}}(q_a, t)$ is an
accepting state but $\delta_{\mathcal{R}}(q_b, t)$ is not, or vice versa. The
test set $T$ generated by the W-method is then defined as $T = C_{\mathcal{R}}
(\{\epsilon\} \cup \Sigma \cup \Sigma^2 \cup \dots \cup \Sigma^{k+1})
D_{\mathcal{R}}$, where $k = m - |Q_{\mathcal{R}}|$ and $T_a T_b = \{t_a t_b
\mid t_a\in T_a, t_b\in T_b \}$ for two sets of traces $T_a$ and $T_b$.

The comprehensiveness of $T$ ensures that any erroneous behavior
in the inferred model will affect the accuracy metrics, and the way in which $T$
is generated does not favor specific parts of the model under analysis, leading
to less skewed accuracy results when compared to statistical estimation.
Nevertheless, using $T$ to measure the accuracy of an inferred model has three
major shortcomings.

First, there are multiple approaches in the area of MBT for DFA with different
aims. For example, the Wp-method~\cite{Fujiwara1991wpMethod} is a further
development of the W-method aimed at reducing the size of $T$ while providing
the same guarantees. Since different approaches will generate different sets of
tests $T$, usually with different number of tests accepted or rejected by either
of the models, they will also lead to different accuracy results.

Second, as already noted by \citet{Walkinshaw2008}, MBT methods often generate 
a set $T$ containing a disproportionately large amount of traces not accepted by
the reference model, which may lead to skewed accuracy results. 
Furthermore, $T$ can become exceedingly large if the difference between the 
number of states of $\mathcal{R}$ and the number of states of $\mathcal{H}$ 
is large. For example, in the W-method described above, the cardinality of 
$T$ grows with $|\Sigma^{k+1}|$, where $k$ is the difference between the 
number of states. This is a significant issue especially in the context of 
model inference from positive examples only, where it is not unusual to obtain 
inferred models having a number of states that is one order of magnitude 
larger than the number of states of the reference model, making this type of 
assessment infeasible. This will also be shown experimentally in 
Section~\ref{sec:experimentalEval}.

\begin{figure}[bt]
	\centering
	\begin{subfigure}[b]{0.4\textwidth}
		\centering
 		\includegraphics[width=\textwidth]{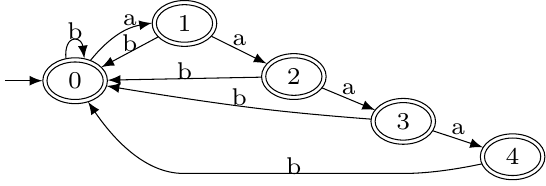}
		\caption{Reference model $\mathcal{R}$}\label{fig:w-method-failure-reference}
	\end{subfigure}
	\hfill
	\begin{subfigure}[b]{0.4\textwidth}
		\centering
 		\includegraphics[width=\textwidth]{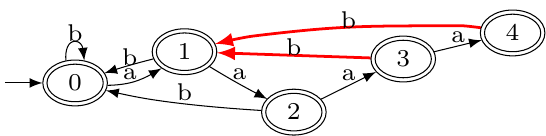}
		\caption{An incorrect model $\mathcal{H}_1$}\label{fig:w-method-failure-h1}
	\end{subfigure}
	\hfill
	\begin{subfigure}[b]{0.4\textwidth}
		\centering
		\includegraphics[width=\textwidth]{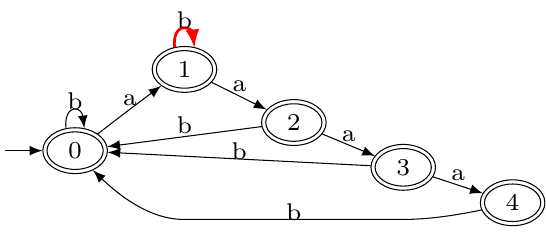}
		\caption{Another incorrect model $\mathcal{H}_2$}\label{fig:w-method-failure-h2}
	\end{subfigure}
    \caption{A example case where the assessment based on the W-method would give misleading results.}\label{fig:w-method-failure}
\end{figure}

Third, although MBT guarantees that the generated $T$ will certainly contain a
counterexample trace highlighting any incorrect behavior of the inferred model,
each counterexample trace has the same ``weight'' on the computed accuracy metrics, even
though it could represent a smaller or larger class of errors. Let us consider
the example shown in Figure~\ref{fig:w-method-failure}; it contains a reference
model $\mathcal{R}$ (Figure~\ref{fig:w-method-failure-reference}) and two
incorrect models $\mathcal{H}_1$ (Figure~\ref{fig:w-method-failure-h1}) and
$\mathcal{H}_2$ (figure~\ref{fig:w-method-failure-h2}), which were manually
created introducing the erroneous transitions highlighted in red. Both
$\mathcal{H}_1$ and $\mathcal{H}_2$ have only \textit{false negative} errors. 
By enumerating all traces we verified that, for any trace length,
$\mathcal{H}_1$ has fewer false negatives than $\mathcal{H}_2$ (e.g., for traces
of length less than or equal to 10, $\mathcal{H}_1$ has 11 false negative
traces, while $\mathcal{H}_2$ has 64 false negative traces). However, when the
model accuracy is evaluated using the set of traces $T$ generated by the
W-method, $\mathcal{H}_1$ has 96\% recall while $\mathcal{H}_2$ has 98\% recall
(precision is 100\% in both cases). This happens because individual false
negative traces in $T$ do not represent an equal number of false negative traces
in the inferred languages.

 \section{Measuring Accuracy with Trace Counting}\label{sec:our_approach} 

To overcome the limitations of statistical and model-based accuracy evaluations, 
we propose to compute precision and recall, 
using analytical measures for the cardinality of
the languages of \textit{true positives}, \textit{false positives} and
\textit{false negatives} --- normally, restricted up to a finite maximum trace 
length to handle languages with unbounded traces. 

In particular, the proposed cardinality measures should be: 
1) \emph{deterministic}, 
making the process repeatable and
avoiding the convergence limitations of 
statistical methods; 
2) \emph{comprehensive}, accounting for every trace;
3) \emph{unbiased}, giving each trace the same weight on the result; 
4) \emph{model-independent}, generating the same accuracy measurement
on all models accepting the same language.

Notably, a statistical estimation in which all the traces (up to the prescribed 
maximum length) are sampled uniformly with the same probability (such as via the 
model-independent random sampling from $\Sigma^*$ 
discussed in Section~\ref{sec:sampling-sigma-star}) would converge to the same 
values we propose to compute analytically. This will be shown experimentally in
Section~\ref{sec:experimentalEval}.

\subsection{Language Cardinality Measures}
We propose to compute two classes of cardinality-based measures to evaluate the 
accuracy of an inferred model (the hypothesis $\mathcal{H}$) against a reference 
model $\mathcal{R}$.

The first step is the construction of the languages corresponding to the definitions 
of \textit{true positive}, \textit{false positive}, and \textit{false negative}, as 
required for the computation of precision and recall. These languages will be accepted,
respectively, by the automata $\mathcal{A}_{TP}$, $\mathcal{A}_{FP}$, and 
$\mathcal{A}_{FN}$ defined as:

\begin{equation}\label{eq:atp-atn-afp-atn}
	\mathcal{A}_{TP} = \mathcal{R} \cap \mathcal{H}; \quad 
	\mathcal{A}_{FP} = \overline{\mathcal{R}} \cap \mathcal{H}; \quad 
	\mathcal{A}_{FN} = \mathcal{R} \cap \overline{\mathcal{H}}
\end{equation}

We can then use analytic combinatorics methods, as described in 
Section~\ref{sec:cardinality_analytic_combinatorics}, to obtain the sequences
$\mathit{tp}_n$, $\mathit{fp}_n$, $\mathit{fn}_n$ of the number of traces of length
$n=0,1,2,\cdots$ accepted by
$\mathcal{A}_{TP}$, $\mathcal{A}_{FP}$, and $\mathcal{A}_{FN}$ respectively --- the 
\emph{cardinality of the languages} accepted by the three automata intersected with 
the evaluation set $E=\Sigma^n$.

Finally, we compose these elementary cardinality measures to compute 
the derived precision and recall metrics for assessing 
an inferred hypothesis model: \emph{cumulative-length} and \emph{single-length}.

\noindent\textbf{Cumulative-Length.} 
This type of assessment considers all the traces of length \emph{up to a
provided value} $n$ (i.e., the evaluation set $E$ is $\Sigma^{\leq n}$). We
compute the cardinalities
\[C_{TP} = \sum_{i=0}^{n} \mathit{tp}_i; \quad  
C_{FP} = \sum_{i=0}^{n} \mathit{fp}_i; \quad C_{FN} = \sum_{i=0}^{n}
\mathit{fn}_i\] 
using the sequences of cardinalities of the languages accepted
by the automata in Equation~\ref{eq:atp-atn-afp-atn}, and then compute
precision and recall:
\begin{equation}\label{eq:precision-recall-with-cardinality-cumulative}
\textit{precision}_{\leq n} = \frac{C_{TP}}{C_{TP} + C_{FP}}; \quad 
\textit{recall}_{\leq n} = \frac{C_{TP}}{C_{TP} + C_{FN}}
\end{equation}

Our analysis does not mandate for a specific value of the maximum trace length
$n$. If any domain knowledge for the application under analysis is available and
suggests the use of a particular value, this should be used. In the absence of
this information, the user should consider that the choice of the maximum trace
length may affect the assessment result --- as we will investigate
experimentally in Section~\ref{sec:rq1_iput_parameter_sensitivity}.

Resolving to using very large values of $n$ so to approximate the
\emph{asymptotic} values of $\textit{precision}_{\leq n}$ and
$\textit{recall}_{\leq n}$ (i.e., the limit of such measures for $n\to\infty$)
is appealing, but should be done with caution, since models with the same
asymptotic values can exhibit different accuracy for shorter trace lengths.
Nonetheless, the asymptotic values of precision and recall carry useful
information. If the precision eventually converges to zero, the hypothesis
language contains more false positives than true positives (in order of
magnitude); if it converges to one, there are more true positives than false
positives, while a value in between these extremes is achievable if the numbers
of true and false positive have the same order of magnitude (including when they
are both finite). Analogous considerations can be formulated for the asymptotic
convergence of the recall measure in
Equation~\eqref{eq:precision-recall-with-cardinality-cumulative} by comparing
the orders of magnitude of the true positives and false negatives languages. We
will further investigate experimentally the behavior of our cardinality-based
accuracy metrics for large values of $n$ in
Section~\ref{sec:rq1_iput_parameter_sensitivity}\footnote{The analytical
computation of such asymptotic values is usually non-trivial and will not be
considered in this work. The interested reader may refer to ``Part B: Complex
Asymptotics'' of~\cite{Flajolet2009} for an extensive treatment of the
subject.}.

\noindent\textbf{Single-length Assessment.} 
A second pair of cardinality-based assessment metrics can be obtained by computing 
precision and recall on the sublanguages of $\mathcal{A}_{TP}$, $\mathcal{A}_{FP}$, and 
$\mathcal{A}_{FN}$ restricted to traces of exactly length $n$, i.e., the traces 
at the intersection of $\Sigma^n$ and the corresponding automaton. 

Given a trace length, computing precision and recall is straightforward,
using directly values from the sequences of cardinalities previously defined:
\begin{equation}\label{eq:precision-recall-with-cardinality-exact}
\textit{precision}_{= n} = \frac{\mathit{tp}_n}{\mathit{tp}_n+ \mathit{fp}_n}; \quad 
\textit{recall}_{= n} = \frac{\mathit{tp}_n}{\mathit{tp}_n + \mathit{fn}_n}
\end{equation}

This assessment type can be used to build a more comprehensive picture of the
model accuracy by repeating the assessment for every trace length within a range
specified by the user. The result has multiple desirable features. First, is not
sensitive to the parameter selection: the choice of the range of trace lengths
over which the single-length assessment is repeated does not affect the result,
but just defines the scope of the assessment. Second, it makes clear how the
model accuracy changes across the trace lengths considered --- a characteristic
that current popular assessment methods do not highlight despite the fact that,
in general, models \emph{do} have variable accuracy depending on the trace
length, as it will be observed in our experimental evaluation
(Section~\ref{sec:experimentalEval}). Third, it allows computing derived
accuracy metrics that consider different trace lengths with different weight.
This may be used, for example, to weight precision and recall on the frequency
of trace lengths observed in a specific deployment of the system.

Although our method does not mandate for a specific range, it should be wide
enough to cover all the possible behaviors of the models under analysis, for
example ensuring that the upper bound of the range is at least equal to the
number of states of the largest automaton under analysis.

 \section{Fast Computation of OGFs for Model Assessment}\label{sec:fastOGF}

In our model assessment method we compute the values of precision and recall
using the counts of \textit{true positives}, \textit{false positives} and
\textit{false negatives}, up to a finite maximum trace length. To obtain these
counts without explicitly enumerating all the possible traces, we use analytic
combinatorics (see Section~\ref{sec:cardinality_analytic_combinatorics}), to
count the number of different accepting paths on the automata accepting these
languages.

To do so, the first step is to obtain the \emph{ordinary generating function}
(OGF) for the cardinality sequence of the language under analysis. This is
generally done using the \emph{transfer matrix
method}~\cite{Flajolet2009,stanley2011enumerativeCombinatorics}, which relies on
linear algebraic operations on the matrix representing the transition relation
of the automaton~\cite{Flajolet2009}, resulting in its worst-case complexity being cubic in the state count. 
This section describes an alternative
state elimination algorithm that, 
despite having the same worst case complexity,
in practice in our preliminary evaluation performed
significantly better than the \textit{transfer matrix method},
by exploiting the sparsity of the transition matrix.

Our method is analogous to the state elimination algorithm by Brzozowski and
McCluskey~\cite{Brzozowski1963fsaToRegex} to generate a regular expression given
a finite state automaton. While in the Brzozowski and McCluskey's algorithm each
transition is labeled with a regular expression indicating the language that
causes traversing the transition, in our method it is labeled with the OGF of
the cardinality sequence of that same language. The reduction rules used when a
state is eliminated then allow us to progressively build the OGF through operations
between rational functions. 

Our method is presented in algorithm~\ref{alg:fastOgf} 
\begin{algorithm}[htb]
	\SetAlgoLined
	\DontPrintSemicolon 
	\caption{Fast computation of the OGF of the sequence of cardinalities of the
	language accepted by a DFA.}\label{alg:fastOgf} 
	\KwData{DFA $\mathcal{A} = (\Sigma, Q,
	q_{i}, F, \delta)$} 
	\KwResult{The OGF of the sequence of cardinalities of the language accepted
	by $\mathcal{A}$} 
	$(\mathcal{G} = \langle N, E \rangle, \mathit{initial},
	\mathit{final}) \leftarrow
	\texttt{digraphConstruction}(\mathcal{A})$\;\label{loc:fastOgf_digraphConstr}
	\While{$\texttt{hasNonInitialOrFinalNode}(\mathcal{G}, \mathit{initial},
	\mathit{final})$}{\label{loc:fastOgf_loopBegin}
		$n \leftarrow \texttt{chooseNonInitialOrFinalNode}(\mathcal{G},
		\mathit{initial}, \mathit{final})$\;
		$\texttt{eliminateNode}(\mathcal{G}, n)$\;
	}\label{loc:fastOgf_loopEnd}
	\Return $E(\mathit{initial}, \mathit{final}\,)$
\end{algorithm}
and works as follows.

\begin{figure}[hbt]
	\centering
	\begin{subfigure}[b]{0.35\textwidth}
		\centering
		\includegraphics[width=\textwidth]{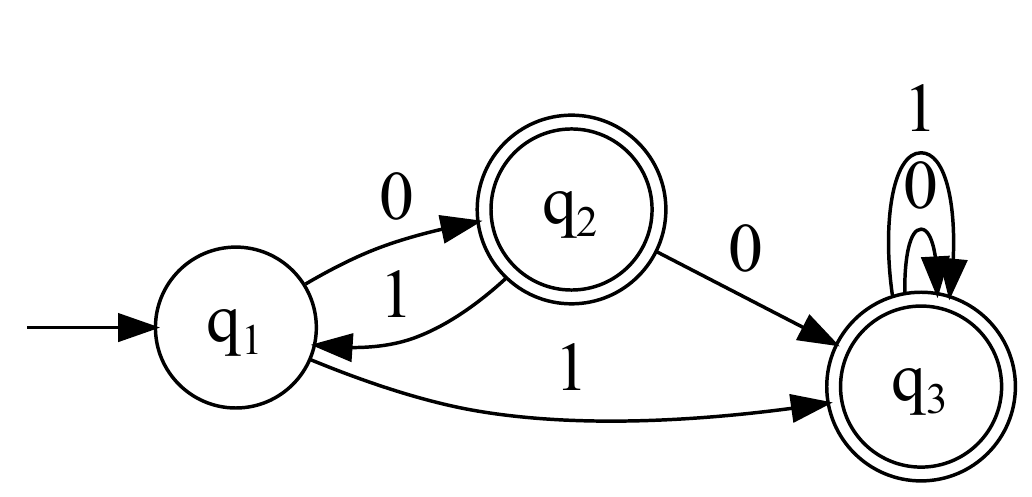}
		\caption{Initial DFA}\label{fig:exampleDfa}
	\end{subfigure}\hfill
	\begin{subfigure}[b]{0.5\textwidth}
		\centering
		\includegraphics[width=\textwidth]{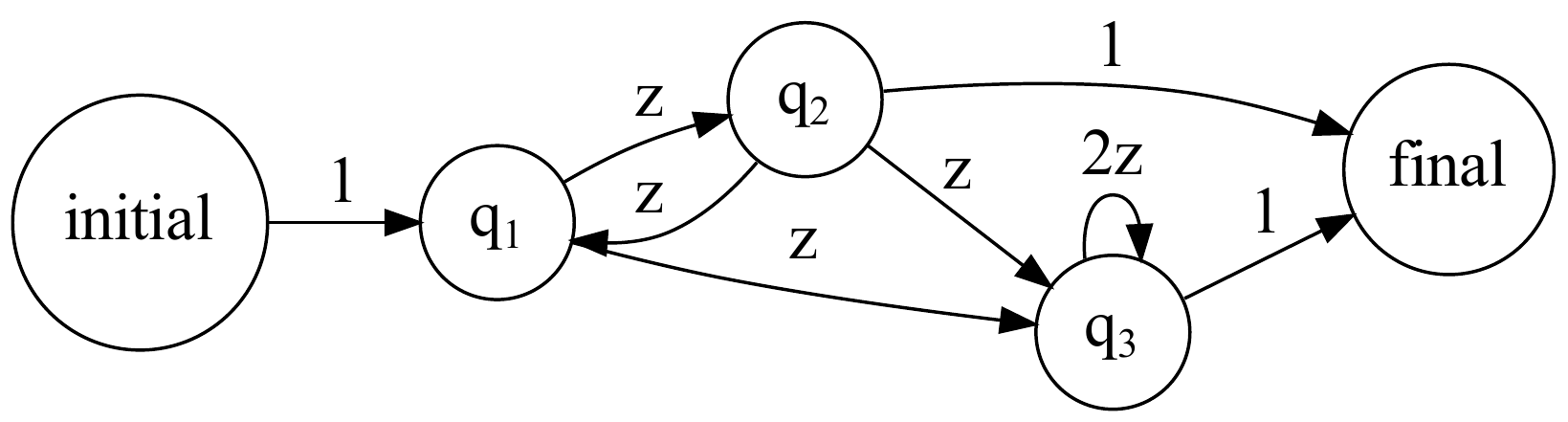}
		\caption{Corresponding digraph $\mathcal{G}$}\label{fig:exampleDfaToG}
	\end{subfigure}
	\caption{Example of digraph construction}\label{fig:constructionOfG}
\end{figure}

Given an input DFA, we construct (line~\ref{loc:fastOgf_digraphConstr}) a directed graph $\mathcal{G}$, using
\texttt{digraphConstruction} (algorithm~\ref{alg:digraphConstr}): 
\begin{algorithm}[tb]
	\SetAlgoLined
	\DontPrintSemicolon 
	\caption{\texttt{digraphConstruction}. Generates the digraph corresponding
	to the given DFA.}\label{alg:digraphConstr} 
	\KwData{$\mathcal{A} = (\Sigma, Q, q_{i}, F, \delta)$ deterministic
	finite-state automaton over the alphabet $\Sigma$.}
	\KwResult{$\mathcal{G} = \langle N, E \rangle$ digraph with edges labeled
	with rational functions, where $N$ is the set of nodes, and $E : N \times N
	\rightarrow RF$ are the edges labeled with rational functions (0 if no edge
	is present).} 
	$N \leftarrow Q \cup \{\mathit{initial}, \mathit{final}\}$\;
	\ForEach{$(a, b) \in N \times N$}{
		$E(a, b) \leftarrow 0$\;	
	}
	\ForEach{$q \in Q$}{
		\ForEach{$s \in \Sigma$}{
			$t \leftarrow \delta(q, s)$\;
			$E(q, t) \leftarrow E(q, t) + z$\;
		}
	}
	$E(\mathit{initial}, q_i) \leftarrow 1$\;
	\ForEach{$q \in F$}{
		$E(q, \mathit{final}) \leftarrow 1$\;
	}
	$\mathcal{G} = \langle N, E \rangle$\; \Return $(\mathcal{G},
	\mathit{initial}, \mathit{final}\,)$
\end{algorithm}
each state of the DFA corresponds to one
node of $\mathcal{G}$ (distinct states correspond to distinct nodes), and the
edge between any ordered pair of nodes of $\mathcal{G}$ is labeled with the
generating function of the sequence of cardinalities of the language of words of
length one, containing the symbols causing the transition between the two
corresponding states of the DFA. If the cardinality of this
language is $n$ (i.e., there are $n$ symbols causing the transition from the
source state to the destination state), the cardinality sequence is $\{0, n, 0,
0, 0, \dots\}$, thus the generating function on the corresponding edge of
$\mathcal{G}$ is $\mathfrak{G}(z) = nz$. In addition, we add to $\mathcal{G}$
a node called \textit{initial}, an edge from \textit{initial} to the node
corresponding to the initial state of the DFA, a node called \textit{final},
and an edge from each node corresponding to an accepting state of the DFA to
\textit{final}. All these additional edges are labeled with the generating
function $\mathfrak{G}(z) = 1$, which is the generating function of the
cardinality sequence $\{1, 0, 0, \dots\}$ (i.e., of the language containing
only the empty string). Adding these edges does not change the final generating
function, but it is a simple way of dealing with multiple final states and
transitions back to the initial state.
Figure~\ref{fig:exampleDfa} 
shows the example DFA taken from~\cite{Aydin2015}
(which computes the OGF using the transfer matrix method), whereas
Figure~\ref{fig:exampleDfaToG} depicts the graph~$\mathcal{G}$ constructed with these
rules described above. 

\begin{figure}[htb]
	\centering
	\hfill
	\begin{subfigure}[b]{0.43\textwidth}
		\centering
		\includegraphics[width=\textwidth]{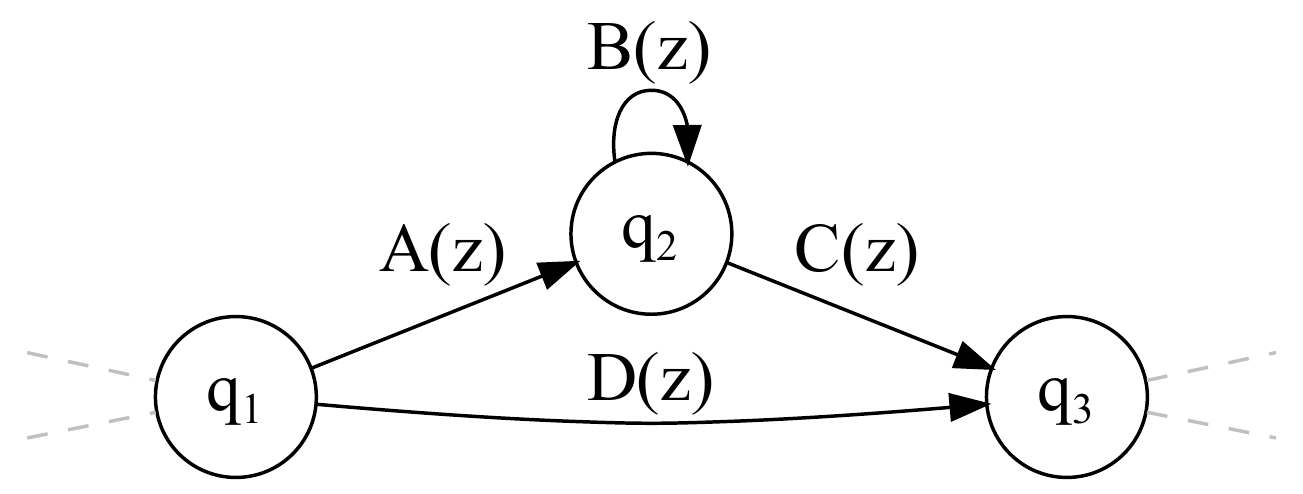}
		\caption{Before eliminating $q_2$}\label{fig:nodeElimination_before}
	\end{subfigure}
	\hfill
	\begin{subfigure}[b]{0.42\textwidth}
		\centering
		\includegraphics[width=\textwidth]{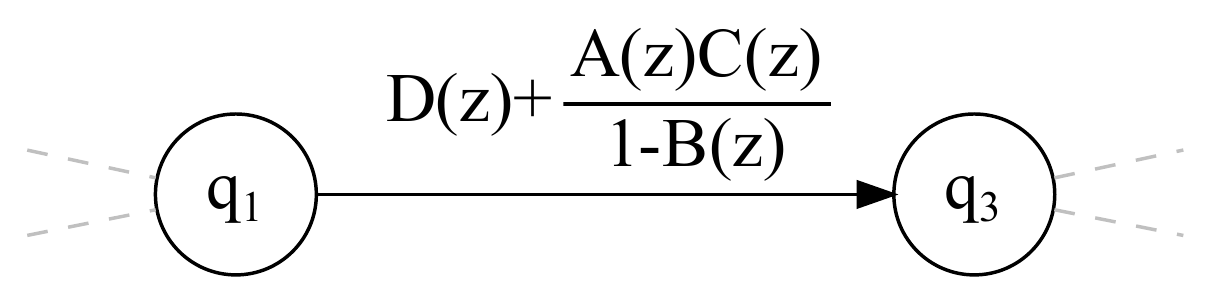}
		\caption{After eliminating $q_2$}\label{fig:nodeElimination_after}
	\end{subfigure}
	\hfill
	\caption{Node elimination}\label{fig:nodeElimination}
\end{figure}

Then, at lines~\ref{loc:fastOgf_loopBegin}--\ref{loc:fastOgf_loopEnd} of
algorithm~\ref{alg:fastOgf}, we eliminate, one by one and using \texttt{eliminateNode}
(Algorithm~\ref{alg:stateElimination}), all the nodes of $\mathcal{G}$ except
\textit{initial} and \textit{final}, as follows.
\begin{algorithm}[tbh]
	\caption{\texttt{eliminateNode}. Eliminates a node while maintaining the correct OGF on the  
	remaining digraph edges.}\label{alg:stateElimination}
	\SetAlgoLined
	\DontPrintSemicolon
	\KwData{$\mathcal{G} = \langle N, E \rangle$ labeled directed graph, where
	$N$ is the set of nodes, and $E : N \times N \rightarrow RF$ are the edges labeled
	with rational functions (0 if no edge is present); \break $n \in N$ the node to be eliminated.}
	\KwResult{The node $n$ is removed from $\mathcal{G}$ and the OGFs on the remaining edges are updated.}
	
	$\mathfrak{G}_{\mathit{loop}} \leftarrow \frac{1}{1 - E(n,
	n)}$\;\label{loc:stateElimination_loopOgf} 
	$E(n, n) \leftarrow 0$ \tcc*[r]{remove self loop if present} 

	$P \leftarrow \{ p \in N : E(p, n) \neq 0 \}$ \tcc*[r]{predecessors of $n$} 
	$S \leftarrow \{ s \in N : E(n, s) \neq 0 \}$ \tcc*[r]{successors of $n$}

	\ForEach{$p \in P$}{
		\ForEach{$s \in S$}{
			$E(p, s) \leftarrow E(p, s) + E(p, n) \,E(n, s) \, \mathfrak{G}_{\mathit{loop}}$\;		
		}
		$E(p, n) \leftarrow 0$\;
	}
	\ForEach{$s \in S$}{
		$E(n, s) \leftarrow 0$\;		
	}
	$N \leftarrow N \setminus n$
\end{algorithm}
Let $n$ be the node to be eliminated; algorithm \texttt{eliminateNode} works
by replacing every pair composed by one edge entering $n$ from a predecessor
node $p$ and one edge exiting $n$ to a successor node $s$, with a single edge
from $p$ to $s$, keeping into account the loop edge from $n$ to $n$ if present.
The procedure computes the OGF on the new edges using the
fact that union, concatenation, and Kleene star of regular languages translate
to algebraic operations between the OGF of the cardinality sequences of the operand
languages. Specifically: if $A$ and $B$ are regular languages, their
\textbf{concatenation} $C = A \cdot B$ is a regular language with
$\mathfrak{G}_C(z) = \mathfrak{G}_A(z) \mathfrak{G}_B(z)$; if $A$ and $B$ are
regular languages and $A \cap B = \emptyset$ (which in our case is a consequence
of the automaton determinism), then their \textbf{union} $C = A \cup B$ is a
regular language with $\mathfrak{G}_C(z) = \mathfrak{G}_A(z) +
\mathfrak{G}_B(z)$; if $A$ is a regular language with $\mathfrak{G}_A(z) = A(z)$
then $A^*$ (\textbf{Kleene star}) is a regular language with
$\mathfrak{G}_{A^*}(z) = \frac{1}{1 - A(z)}$. 
As a result, Algorithm~\ref{alg:stateElimination} computes the OGF of the Kleene
star of the language that would be on the loop of the node to be eliminated
(line~\ref{loc:stateElimination_loopOgf}), eliminates the loop (line 2) ---
otherwise the set of predecessors and successors subsequently defined would
contain $n$ --- and then iterates through the edges entering $n$ and exiting $n$,
updating the OGF on the edge from the predecessor node to the successor node
as illustrated in Figure~\ref{fig:nodeElimination}.
When \textit{initial} and \textit{final} are the only nodes left in
$\mathcal{G}$, the OGF on the edge from \textit{initial} to \textit{final} is
the OGF of the cardinality sequence of the language accepted by the DFA.

\begin{figure}[tbh]
	\centering
	\begin{subfigure}[b]{0.39\textwidth}
		\centering
		\includegraphics[width=\textwidth]{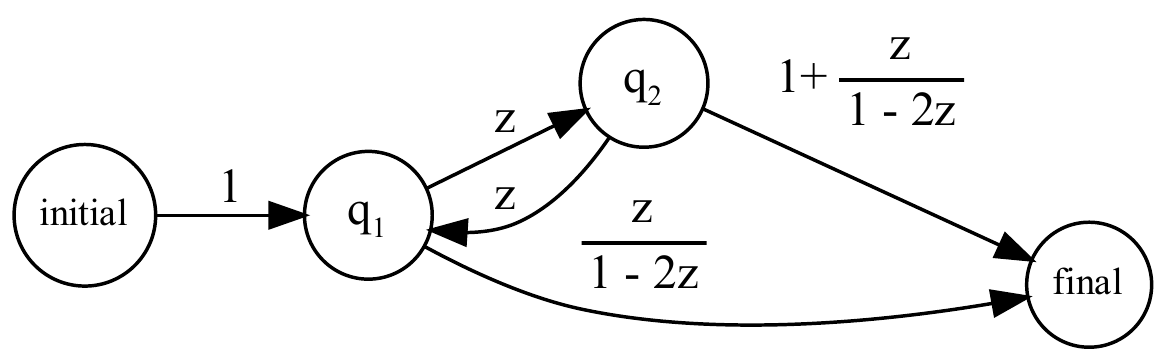}
		\caption{After eliminating $q_3$}\label{fig:exampleDfaElimQ3}
	\end{subfigure}
	\hfill
	\begin{subfigure}[b]{0.29\textwidth}
		\centering
		\includegraphics[width=\textwidth]{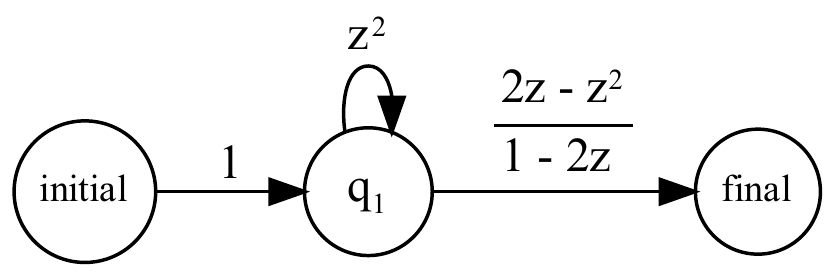}
		\caption{After eliminating $q_2$}\label{fig:exampleDfaElimQ2}
	\end{subfigure}
	\hfill
	\begin{subfigure}[b]{0.29\textwidth}
		\centering
		\includegraphics[width=\textwidth]{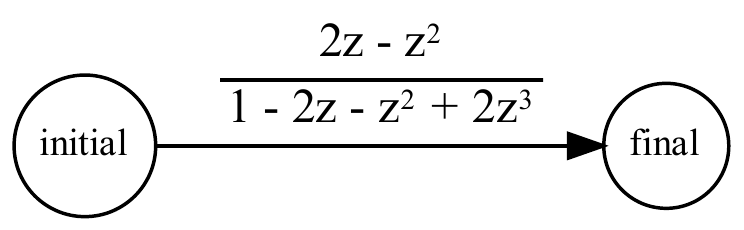}
		\caption{After eliminating $q_1$}\label{fig:exampleDfaElimQ1}
	\end{subfigure}
	   \caption{Example sequence of state eliminations for  the DFA shown in Figure~\ref{fig:exampleDfa}}\label{fig:eliminationExampleDfa}
\end{figure}

Figure~\ref{fig:eliminationExampleDfa} shows one possible sequence of state 
eliminations to obtain the generating functions of the DFA shown in Figure~\ref{fig:exampleDfa}.

We remark the order with which the nodes are eliminated is not relevant for the
correctness of the result. However, it does significantly affect the performance
due to the cost of the operations between rational functions required to compute
the generating functions on the edges involved in each elimination. A simple
heuristic that proved to be beneficial in our experiments is choosing first the
node with the lowest OGF degree\footnote{The OGF degree is defined as the maximum of the degrees
of its constituent numerator and denominator polynomials.} on the self loop (thus
the nodes with no self loop are eliminated first).

 \section{Experimental Evaluation}\label{sec:experimentalEval} In this section,
we report on the experimental evaluation of our model assessment method in terms of the following
aspects.
First, since the cumulative-length assessment requires, as an input parameter, a
maximum trace length, we investigate the impact of changing its value on the
model assessment results. We also analyze whether the asymptotic values (i.e.,
the results obtained when the maximum trace length go to infinity) are useful to
characterize the model accuracy.
Second, to better understand to what extent the model assessment results differ
depending on the methods used, we compare the results obtained using our method
with the results obtained using existing methods, specifically the statistical
estimation and MBT-based methods.
Third, in Section~\ref{sec:our_approach}, we claimed that a
statistical estimation in which traces having the same length have the same
probability of being sampled would converge to the same results obtained by our
single-length assessment method. We will verify this experimentally by comparing
the single-length assessment and a model-independent sampling of $\Sigma^*$
in terms of the precision and recall values.
Finally, we evaluate the applicability of our method
on models inferred using well known model inference approaches.

To summarize, we address the following research questions:
\begin{enumerate}[\bf RQ1:]
	\item \textit{How does the choice of the maximum 
    trace length affect the model assessment results?}
	\item \textit{How do the values of precision and recall 
    obtained using our method compare with the results obtained using 
    other evaluation methods?}
    \item \textit{Is the single-length assessment over a range of lengths a
    usable alternative to the statistical evaluation of model accuracy using
    model-independent sampling of $\Sigma^*$?}
    \item \textit{Is our method applicable to the assessment of inferred models 
    representing aspects of actual software systems?} 
\end{enumerate}

\subsection{Evaluation Subjects}\label{sec:experimentalEval_subjects}
To evaluate model assessment methods, we need various pairs of reference and
inferred models. We describe how we select reference models and generate
inferred models from them below.

\subsubsection*{Reference Models}

We selected 41 publicly available reference models, taken from existing
studies~\cite{Pradel2010,Krka2014}; all of them have a well-documented origin
and were previously used in the model inference literature. 

\citet{Pradel2010} selected 32 commonly used classes from the Java SDK API 
and identified their method ordering constraints using the API documentation 
and well-known reference books. These constraints were then translated into 
reference models representing all possible valid traces of method calls. 
The resulting reference models are publicly available on 
the authors' website\footnote{\url{http://mp.binaervarianz.de/icsm2010/index.html}}.
These models were used also in previous work on model inference~\cite{Busany2019}.

\citet{Krka2014} selected 9 open-source libraries, found the corresponding
reference models (manually specified in previous work), and checked them
manually for inconsistencies. These models were checked against execution traces
collected from actual executions of software using those libraries, and the
transitions on methods that were never invoked in the collected traces were
eliminated. The resulting reference models are publicly available on the
authors' website\footnote{\url{https://softarch.usc.edu/wiki/doku.php?id=inference:start}}.
These models have been used also in previous work~\cite{Walkinshaw2016, Le2015} 

\subsubsection*{Inferred Models}
Ideally, the inferred models should be generated using model inference engines
on traces produced by executions of actual software systems. Unfortunately, we
were not able to find and execute the exact versions of the software systems
represented by the 41 reference models, to collect their execution traces. As an
alternative, we used the same type of random walk described in
Section~\ref{sec:evaluation_with_trace_similarity} to generate a set of random
traces for each reference model, and then we processed each set of traces using
different model inference engines. 
As for the random walk parameters, we used
the termination probability $p_a=0.1$ and the uniform probability for selecting
an outgoing transition among available transitions following
\citet{Walkinshaw2013stamina}. For each reference model, we repeated the random walk
until at least 100 traces were generated \emph{and} each state of the
model had been visited at least four times, as suggested by \citet{Busany2019}. From each
set of traces we inferred two models, using two different model inference
algorithms: $k$-tails~\cite{biermann1972synthesis} (the most well-known model inference
algorithm) and MINT~\cite{Walkinshaw2016} (a state-of-the-art model
inference technique). 
Before the accuracy assessment, the inferred models were minimized, through standard
(language preserving) automata minimization.

It is worth noting that, in the area of model inference, it is common practice
to use randomly generated traces when traces coming from actual software
executions are not available (e.g.,
~\cite{Busany2019,Lo2012,Walkinshaw2013,Walkinshaw2013stamina}). 
Moreover, model
assessment is independent of how inferred models are generated, thus is not a
major issue in our evaluation. Nevertheless, we will discuss this as a potential
threat to validity in Section~\ref{sec:threats-to-validity}. 

To summarize, our experimental evaluation is based on 82 test subjects (pairs of
reference and inferred models): 41 reference models, with two inferred models
each. Table~\ref{tab:model-characteristics} shows the size of the reference and
inferred models in terms of the minimum, the average, and the maximum number of
states and transitions. 

\begin{table}[htb]
    \caption{Characteristics of reference and inferred models}\label{tab:model-characteristics}
    \begin{tabular}{cccccccc}
    \toprule
        & \multicolumn{3}{c}{State count} &  & \multicolumn{3}{c}{Transition count} \\
        \cmidrule(r){2-4} \cmidrule(r){6-8}
        & Min & Avg  & Max & & Min & Avg & Max \\ \midrule
    Reference & 2 & 9.2 & 41 &  & 7 & 77.8 & 465 \\
    Inferred & 2 & 349 & 2060 & & 5 & 752 & 4994 \\ \bottomrule
    \end{tabular}	
\end{table}

\subsection{Evaluation Settings}
We performed the assessment of all the 82 test subjects 
using our implementation of trace similarity, MBT-based assessment, 
and our method with trace counting, 
all developed in Java and publicly available (see Section~\ref{sec:data-availale}).

All the experiments were executed on an AMD EPYC 
7401P (24 cores, 48 threads) with 448 GB of RAM. Since the implementation 
of our evaluation method is single-thread, 24 evaluations were executed 
concurrently.

The evaluation runtime of every model assessment performed in our 
evaluation was subject to a timeout of two days. It is worth pointing out that 
all the single-length and cumulative-length assessments on the same test 
subject require the same evaluation runtime. The reason is that this evaluation 
runtime is dominated by the time required to compute the generating functions, 
which are the same for all assessments of the same test subject.

\subsection{Data availability}\label{sec:data-availale} To support open science
and enhance the reproducibility of our evaluation, we provide a replication
package\footnote{\url{https://doi.org/10.6084/m9.figshare.24932799}}
including all the artifacts: the source code, the reference models and the
inferred models used.

\subsection{RQ1: Input Parameter Sensitivity}\label{sec:rq1_iput_parameter_sensitivity}

\paragraph{Methodology}

As described in Section~\ref{sec:our_approach}, our method can be applied in two
ways: \emph{cumulative-length} and \emph{single-length} assessments. The
single-length assessment repeated over a range of trace lengths
is not sensitive to parameter changes: the \emph{range} changes just the \emph{scope} of the assessment, i.e.,
for which lengths the values of precision and recall are computed, without
affecting the output values themselves. To evaluate the parameter sensitivity of
the cumulative-length assessment, we computed the precision and recall values
varying the only parameter (i.e., the maximum trace length) from 0 to 200 in
steps of 1.

\paragraph{Results}

\begin{figure}[tb]
  \subfloat[Precision]{
  \includegraphics[height=0.265\textheight]{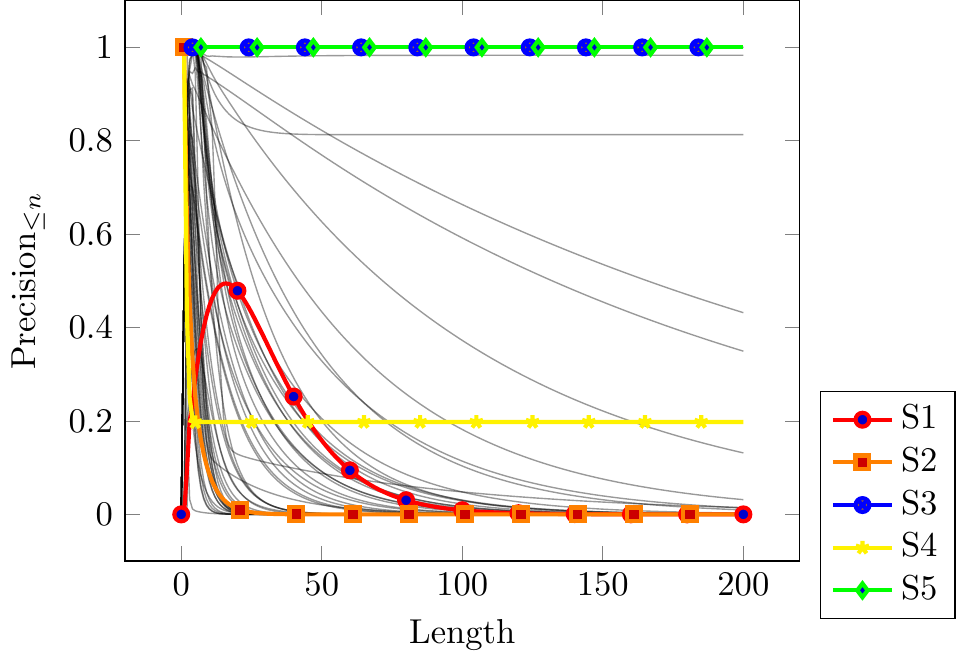}}
  \hspace{-5mm}
  \subfloat[Recall]{\includegraphics[height=0.265\textheight]{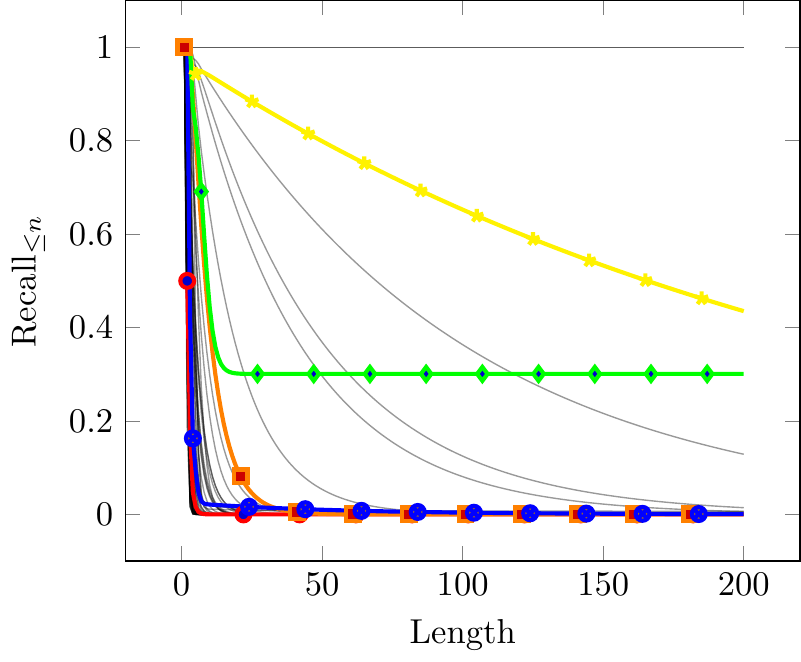}}
  \caption{Values of $\textit{precision}_{\leq n}$ and
  $\textit{recall}_{\leq n}$ measured using the cumulative-length assessment, varying the
  maximum trace length parameter $n$ in the range [0, 200]. One plot per test
  subject. Five test subjects are highlighted, and their characteristics are summarized in Table~\ref{tab:test-subjects-info}.}\label{fig:precision-recall-all}
\end{figure}

All the assessments of 61 of the 82 test subjects terminated within the 2-day
timeout. Figure~\ref{fig:precision-recall-all} shows how the precision and
recall values vary depending on the maximum trace length parameter; each line is
a result for one test subject. 

The foremost takeaway of the evaluation results is that the choice of the
maximum trace length parameter does indeed affect the computed accuracy values,
and it does so differently, depending on the inferred model. 
The results highlighted the prevalence of cases in which precision and/or
recall values cover a large part of the range $[0,1]$, depending on the parameter
value. Specifically, among the 61 test subjects that finished within the
timeout, 20 have a precision value (and 54 have a recall value) that covers the entire range
$[0,1]$, depending on the maximum trace length parameter. This emphasizes
how --- for any choice of a single value of the maximum trace length --- some relevant
information about the model accuracy is inevitably lost, leading to possibly
misleading results.

Figure~\ref{fig:precision-recall-all} also shows the variety of trends that the
precision and recall values can have, as the maximum trace length is
increased: e.g., convergence to 0 with and without a peak, constant equal to 1, and
convergence to $0 < k < 1$. In the figure we have highlighted (with
colors and special markers) five test subjects
that, together, cover the most common trends we observed in the results of this
experiment. Appendix~\ref{sec:appendix_test_subj_selection} gives further details 
on how we selected these test subjects, 
and Appendix~\ref{sec:selected_subjects_descr} describes the selected test subjects in more detail.
Table~\ref{tab:test-subjects-info} summarizes the characteristics of these representative test subjects.

\begin{table}[tbh]
\caption{Characteristics of the selected test subjects}\label{tab:test-subjects-info}
    \begin{tabular}{@{}ccccccc@{}}
    \toprule
        & \multicolumn{3}{c}{Reference model} & \multirow{2}{*}{\begin{tabular}[c]{@{}c@{}}Inference\\
        approach\end{tabular}} & \multicolumn{2}{c}{Inferred model} \\
        \cmidrule(r){2-4} \cmidrule(r){6-7}
        & Name & States  & Trans. & & States & Trans. \\ \midrule
    Subject 1 (S1) & java.net.URL \cite{Pradel2010} & 5 & 57 & MINT & 718 & 993 \\
    Subject 2 (S2) & SMTPProtocol \cite{Krka2014}  & 3 & 14 & MINT & 24 & 110 \\
    Subject 3 (S3) & StringTokenizer \cite{Pradel2010}  & 4 & 14 & 2-tails & 83 & 175 \\
    Subject 4 (S4) & Signature \cite{Krka2014} & 3 & 7 & MINT & 15 & 63 \\
    Subject 5 (S5) & StringTokenizer \cite{Krka2014}  & 4 & 7 & 2-tails & 30 & 61 \\ \bottomrule
    \end{tabular}	
\end{table}

Furthermore, Figure~\ref{fig:precision-recall-all} shows that the asymptotic 
value\footnote{
  We determined the asymptotic value by manually checking to which value the
  accuracy stabilizes. There were some cases (seven for the precision value,
  four for recall value) in which we were not able to determine the asymptotic
  value because, within the observed range, it was not clear around which value
  it stabilized.
  } can be reached 
through different paths, with fast or slow convergence, and with or without
local minima or maxima. 
This confirms that judging or comparing model accuracies according to the 
asymptotic value of precision and recall can be misleading, since models 
with the same asymptotic accuracy generally can have different accuracies 
for shorter traces. 

Although investigating the asymptotic accuracy of models obtained with current
inference methods is outside the scope of this paper, an interesting 
observation is that the asymptotic values of the accuracy metrics are 
generally either zero or one. This is because the precision and recall values are 
quotients between language cardinalities that tend to infinity as the maximum 
trace length considered is increased, therefore the quotient often goes to zero
or one depending on the relative growth  rate of the languages in
Equation~\ref{eq:precision-recall-background}.
In our evaluation, the most common asymptotic value is zero: 45 subjects have a 
precision value converging to zero, while 55 have a recall value converging to zero. 
The number of subjects having accuracy converging to a value greater than zero 
and smaller than one is only three for precision and two for recall. This is an 
observation that is outside the scope of this paper, and may not generalize to 
models obtained with different inference methods; it will be further investigated 
in future work.

It is difficult to relate accuracy trend patterns (observed a posteriori)
with characteristics of the inferred model such as the topology. However, it is
worth noting that inferred models accepting finite languages (identified by the
absence of loops) generally indicate that the inference algorithm was not able
to generalize the training data, resulting in inferred models with good
precision but poor recall. Some patterns that we observed a posteriori are
discussed in Appendix~\ref{sec:selected_subjects_descr}. It is also worth noting
that the analysis of the generating function associated to the models'
cardinalities can predict the asymptotic trends \cite{Flajolet2009}; nevertheless,
this type of analysis is beyond the scope of this paper.

\subsubsection*{Answer to RQ1}
The maximum trace length parameter does affect the computed
cumulative-length precision and recall values, and it does so differently,
depending on the inferred model. This parameter should be tuned, case by
case, to a value that is relevant for the application under analysis. If no
domain knowledge is available, it is preferable to consider how precision and
recall values change considering different trace lengths, using the single-length
metrics.

\subsection{RQ2: Comparison of Model Assessment Methods}\label{sec:rq2}

This section discusses how the results generated by the assessment methods
described and discussed in the paper compare with each other. In particular, we
will examine trace similarity, MBT-based assessment, cumulative-length
assessment, and single-length assessment. 

Note that the model assessment methods discussed in this paper can be classified
into two types depending on their output types: trace similarity, MBT-based
assessment, and the cumulative-length assessment are all methods that return one
pair of precision and recall values, while the single-length assessment instead
returns one pair of precision and recall values \emph{per trace length} in the
given range (which is a parameter of the method). These two types cannot be
directly compared. For this reason the comparison proposed in this section is
divided in two parts.

In the first part, we will compare trace similarity, MBT-based, and
cumulative-length, with the goal of understanding the consistency and the
differences between the computed precision and recall values across the
different assessment methods and different parameter values.

In the second part, we will discuss the single-length assessment results in
relation with the results generated with the other currently used
methods. In this case a direct comparison is neither possible nor meaningful,
due to the different output type. To make them comparable, we will
\emph{condition the other assessment methods on the trace length}: the
evaluation sets generated by these methods will be partitioned according to the
trace length, and the subsets will be used to compute one pair of precision and
recall values per trace length, which can then be directly compared with the
results from the single-length assessment. 

\subsubsection*{Comparison of Trace Similarity, MBT-based, and 
Cumulative-Length Assessments}\label{sec:comparison-cumulative} 
As discussed above, we first compare trace similarity, MBT-based, and
cumulative-length.

\paragraph{Methodology}
To understand the consistency and the differences between the computed precision
and recall values across the different assessment methods, 
we run the methods on the 82 test subjects, with the following parameters settings.

Trace similarity requires defining how to make the nondeterministic choices
needed to conduct the random walk (i.e., whether to terminate the walk, and
which available transition should be traversed --- see
Section~\ref{sec:evaluation_with_trace_similarity}). This has a direct effect on
the sampling bias, thus affecting the assessment results. In our setup, the
selection of the outgoing transition to be traversed was done uniformly among
the available alternatives, as it is often done in practice
(e.g.,~\cite{Walkinshaw2013stamina, Lo2006}). Regarding the termination of the
trace when a final state is reached, we found different approaches in the model
inference literature. Sometimes the termination probability is a function of the
outdegree of the accepting state (e.g., as in
reference~\cite{Walkinshaw2013stamina}), sometimes is not (e.g., as in
reference~\cite{Lo2012}). To highlight how this choice may affect the assessment
result, we used a fixed termination probability $p_a$, and we executed the trace
similarity assessment setting $p_a$ to three values (0.02, 0.1 and 0.5), thus
obtaining three pairs of precision and recall values for each test subject.
Trace similarity also requires specifying the target size for the evaluation set
$E$, which in our experiments was set to 100,000 traces. To further ensure
adequate model coverage, we continued adding traces to $E$ beyond the target
size, until each transition of the model was followed at least 10 times (as
proposed in reference~\cite{Lo2006}), or a 30-minute time limit was reached. As
a result, in some cases we generated more than 100,000 traces.

The MBT-based assessment requires a model-based testing method for finite state
automata, to generate the evaluation set $E$. We use the W-method, as done by
the original proponents of the method. 

The cumulative-length assessment requires a parameter indicating the maximum
length of the traces to be considered. Since there is no ``correct'' value to choose
in the absence of domain knowledge for a specific test subject, we
evaluate the results when varying the parameter from 0 to 200 in steps of 1.

\paragraph{Results}

All the model assessments performed using trace similarity returned a result
within the timeout. The cumulative-length assessments terminated within the
timeout for 61 of the 82 test subjects (for all parameter values), while for the
remaining 21 subjects a timeout occurred during the computation of the
generating function. The MBT-based assessment terminated within the timeout only
in two of the 82 cases under analysis. The reason is that the cardinality of the
evaluation set generated with the W-method grows with $|\Sigma^{d+1}|$, where
$d$ is the difference in the number of states between the inferred and reference
model, therefore the method is usable only when the reference and inferred
models have a similar number of states. This is rarely the case in our set of
test subjects: the difference $d$ is on average 247 ($\sigma=265$). Due to the
lack of a sufficient number of results from the MBT-based assessment, we omit
this method in the rest of the comparison.

Before we compare the trace similarity and cumulative-length assessment results,
it is worth focusing on how trace similarity is affected by the termination
probability $p_a$ of the random walk. Table~\ref{tab:length-pa} shows the effect
of $p_a$ on the average length of the traces generated in our experimental
evaluation. These are consistent with an exponential distribution, however we
remind that the relationship between termination probability and distribution of
trace lengths is model dependent, and our experiments lead to this distribution
only because most of the non-error states of the models under analysis are
accepting states.

\begin{table}[ht] 
    \caption{Mean trace length generated by the random walk, 
    depending on the $p_a$ parameter.}\label{tab:length-pa}
    \begin{tabular}{lr}
        \toprule
        $p_a$ & Mean trace length \\ \midrule
        0.5 & 1.99 \\
        0.1 & 11.28 \\
        0.02 & 49.45 \\ \bottomrule
    \end{tabular}
\end{table}

Table~\ref{tab:diff-ts} summarizes the absolute difference between the model
accuracy evaluated using trace similarity with different random walk termination
probability $p_a$. For example the mean absolute difference between the
precision value computed using trace similarity with a random walk with $p_a =
0.5$, and the precision value computed using trace similarity with a random walk
with $p_a = 0.02$, is \SI{36}{\pp} ($\sigma = \SI{29}{\pp}$), with \si{\pp}
indicating percentage points. This difference indicates that the choice of the
random walk parameters values can affect the trace similarity result in a way
that does not allow us to reliably determine the model accuracy. Note that a
larger difference in the $p_a$ parameter leads to a larger difference in the
measured accuracy (e.g., line 1 vs line 3 of Table~\ref{tab:diff-ts}). This is
explained intuitively in terms of the sampling bias discussed in
Section~\ref{sec:evaluation_with_trace_similarity}: a random walk with higher
termination probability $p_a$ will generally produce shorter traces, and we
know, from the RQ1 results in Section~\ref{sec:rq1_iput_parameter_sensitivity},
that in most cases the model accuracy is higher on shorter traces.

\begin{table}[ht]
    \caption{Mean and standard deviation of the absolute difference between the
    model accuracy evaluated using trace similarity with different random walk
    termination probability $p_a$. Values in percentage
    points (\si{\pp}).}\label{tab:diff-ts}
    \begin{tabular}{lcc}
        \toprule
         & Mean & $\sigma$ \\ \midrule
        $|\textit{Precision}_{\textit{TS}_{0.5}} - \textit{Precision}_{\textit{TS}_{0.1}}|$ & \SI{19}{\pp} & \SI{14}{\pp} \\
        $|\textit{Precision}_{\textit{TS}_{0.1}} - \textit{Precision}_{\textit{TS}_{0.02}}|$ & \SI{19}{\pp} & \SI{17}{\pp} \\
        $|\textit{Precision}_{\textit{TS}_{0.5}} - \textit{Precision}_{\textit{TS}_{0.02}}|$ & \SI{36}{\pp} & \SI{29}{\pp} \\
        $|\textit{Recall}_{\textit{TS}_{0.5}} - \textit{Recall}_{\textit{TS}_{0.1}}|$ & \SI{37}{\pp} & \SI{20}{\pp} \\
        $|\textit{Recall}_{\textit{TS}_{0.1}} - \textit{Recall}_{\textit{TS}_{0.02}}|$ & \SI{15}{\pp} & \SI{10}{\pp} \\
        $|\textit{Recall}_{\textit{TS}_{0.5}} - \textit{Recall}_{\textit{TS}_{0.02}}|$ & \SI{52}{\pp} & \SI{27}{\pp} \\ \bottomrule
    \end{tabular}
\end{table}

\begin{figure*}[ht]
	\centering
    \includegraphics[width=\textwidth]{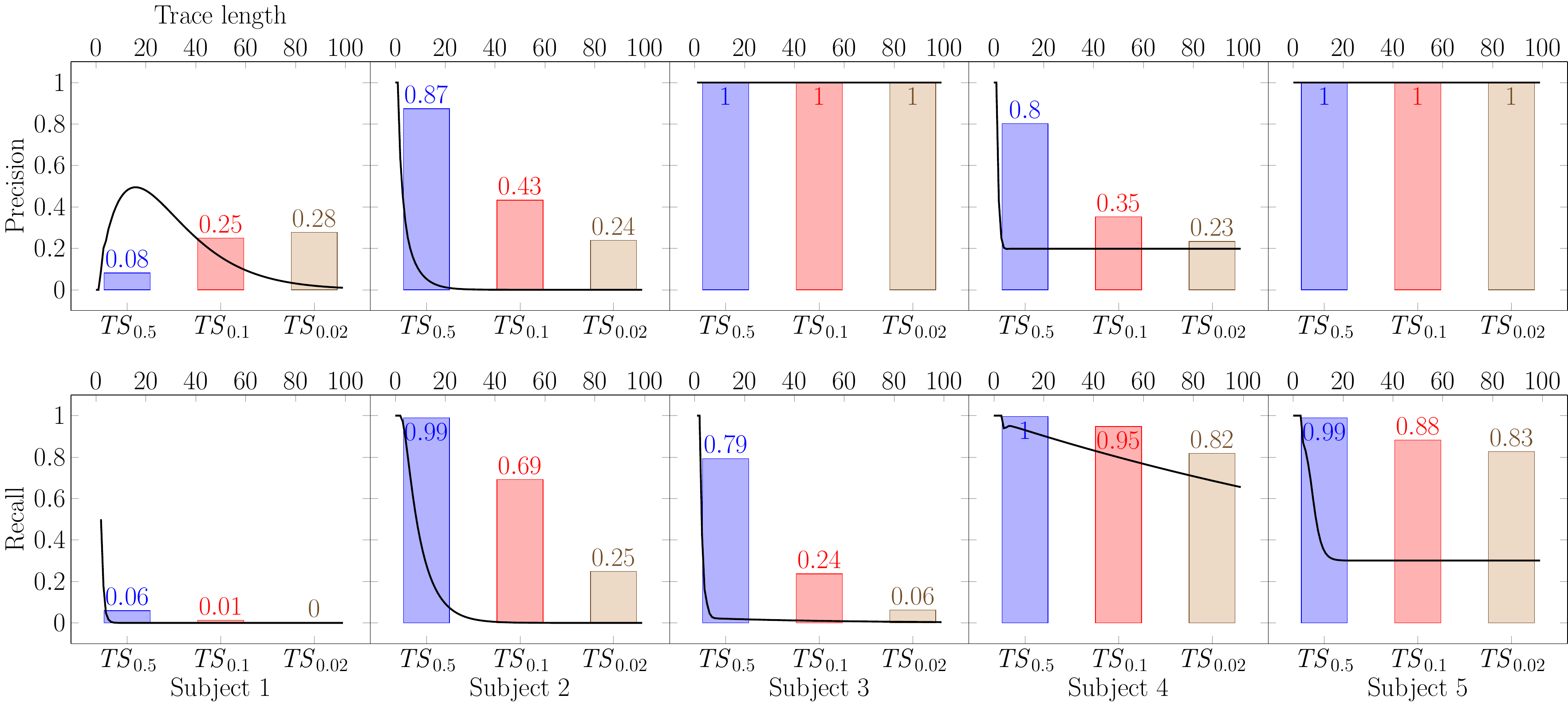}
	\caption{Comparison of trace similarity with three different random walks  
    ($\textit{TS}_\textit{0.5}$, $\textit{TS}_\textit{0.1}$ and $\textit{TS}_\textit{0.02}$), 
    and the cumulative evaluation proposed in  
    Section~\ref{sec:our_approach} (black plot) over five
    representative test subjects.}\label{fig:comparison-cumulative}
\end{figure*}

\figurename~\ref{fig:comparison-cumulative} shows the comparison of the
precision and recall values computed using trace similarity 
and our cumulative-length assessment, for the five representative
test subjects selected in RQ1. To improve the readability, the range of the
trace length parameter for the cumulative-length assessment shown in the figure
is from 0 to 100. The same information is available for all test subjects in the
supporting material (see Section~\ref{sec:data-availale}). Note the three
different trace similarity assessments per test subject (using different values
of the $p_a$ parameter of the random walk).

The most important observation is that both trace similarity and
cumulative-length results values are different in most cases, and vary over a
large range of values depending on the methods' parameters. 

A closer look reveals a consistency between the trend of the accuracy results
computed using trace similarity as $p_a$ is decreased, and the trend of the
accuracy results computed by the cumulative-length assessment for longer traces:
in all subjects for which the cumulative-length assessment shows decreasing
model accuracy for longer traces, also the trace similarity assessment shows
decreasing accuracy as $p_a$ is decreased from 0.5 to
0.02. Analogously, the cumulative-length assessment for subject
S1 shows an (initially) increasing precision, and
the corresponding trace similarity result is also consistent with this trend.
The reason behind this consistency is the effect of the random walk termination
probability $p_a$ on the length of the traces generated by the random walk, as
discussed before. However, it is important to stress that the distribution of
trace lengths generated by the random walk is not enough to fully capture how
the sampling bias affects the trace similarity result, because also different
traces of the same length may have different probabilities of being generated.
The nondeterministic choices required to conduct the random walk can be
performed in a variety of ways, and their effect on the result is
model-dependent. On the other hand, the cumulative-length assessment has only
the maximum trace length parameter, whose effect can be intuitively understood,
making it easier to tune.
Section~\ref{sec:comparison-conditioned} will include a scenario in which trace
similarity and cumulative-length assessment diverge due to this effect. 

Furthermore, in \figurename~\ref{fig:comparison-cumulative} we can observe that
for some test subjects all the model assessment methods compute the same value
(100\%) for precision. We manually verified that in these cases the inferred
language is a subset of the reference language, hence the evaluation set cannot
contain false positive traces, regardless how it is generated, and thus
the computed precision value is always 100\%.

\subsubsection*{Comparison of Statistical Estimation and 
Single-Length Assessment}\label{sec:comparison-conditioned}

We now turn our attention to the results of the single-length assessment over a
range of trace lengths, and how it compares with the results of trace
similarity.

\paragraph{Methodology} 
As discussed above, a direct comparison is not possible, since the single-length
assessment returns one pair of precision and recall values \emph{per trace
length}, while trace similarity returns one pair of precision and recall values
overall. 

To obtain comparable results, we will \emph{condition the trace similarity on
the trace length}, by partitioning each evaluation set $E$ according to the
trace length. This will allow us to use the resulting subsets to compute one
pair of precision and recall values per trace length, and it will also allow us
to look at the distribution of the traces in $E$ across different trace lengths.
In fact, we will initially focus on this distribution, which is the first effect
of the sampling bias induced by the random walk mentioned in
Section~\ref{sec:evaluation_with_trace_similarity}. Then, we will compare the
values of precision and recall for different trace lengths (from 1 to 100 in
steps of 1). Since our single-length assessment has no sampling bias, any
difference in precision or recall value for a specific trace length implies
statistical noise and/or the non-uniformity of the sampling induced by the
random walk (i.e., the second effect of the sampling bias as mentioned in
Section~\ref{sec:evaluation_with_trace_similarity}). 

\paragraph{Results}

\begin{figure}[tbh]
	\centering
	\begin{subfigure}[b]{\textwidth}
		\centering
		\includegraphics[width=0.9\textwidth]{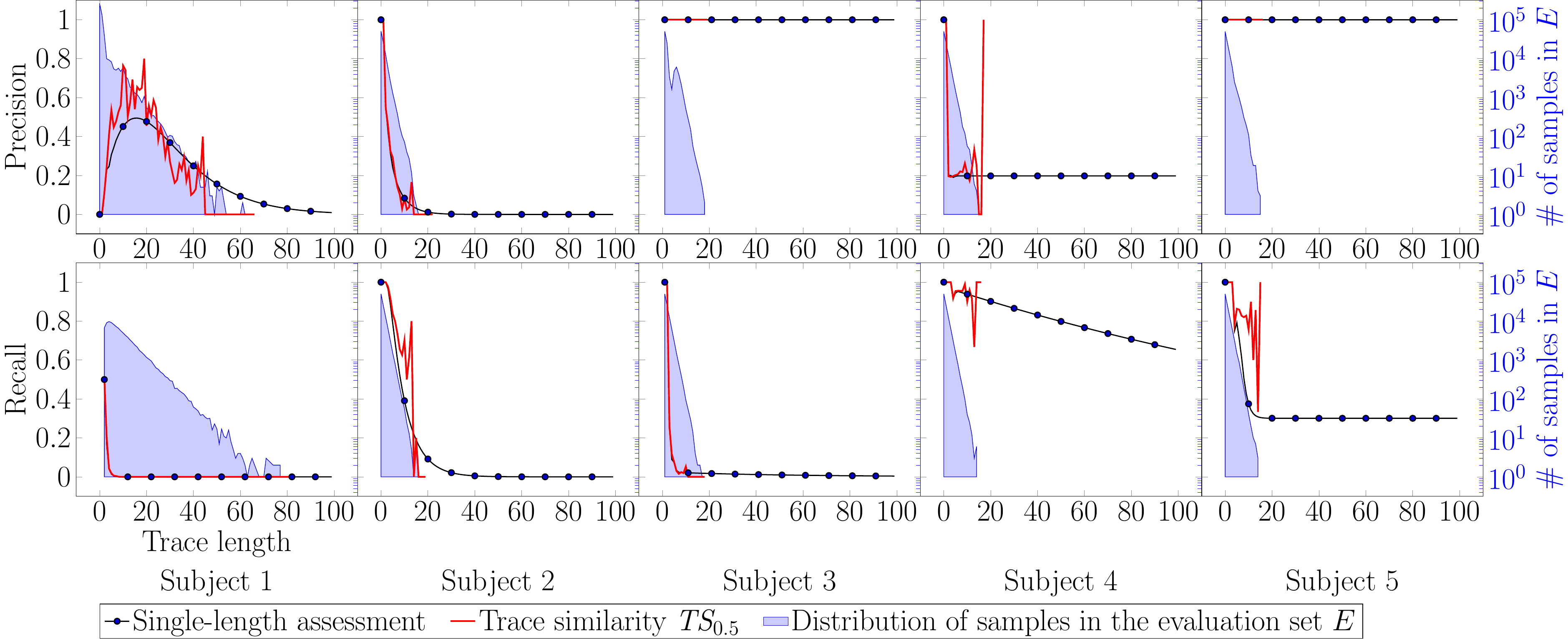}
		\caption{Trace similarity using a random walk with $p_a = 0.5$}\label{fig:comparison-partitioned-p5}
	\end{subfigure}
	\hfill
	\begin{subfigure}[b]{\textwidth}
		\centering
		\includegraphics[width=0.9\textwidth]{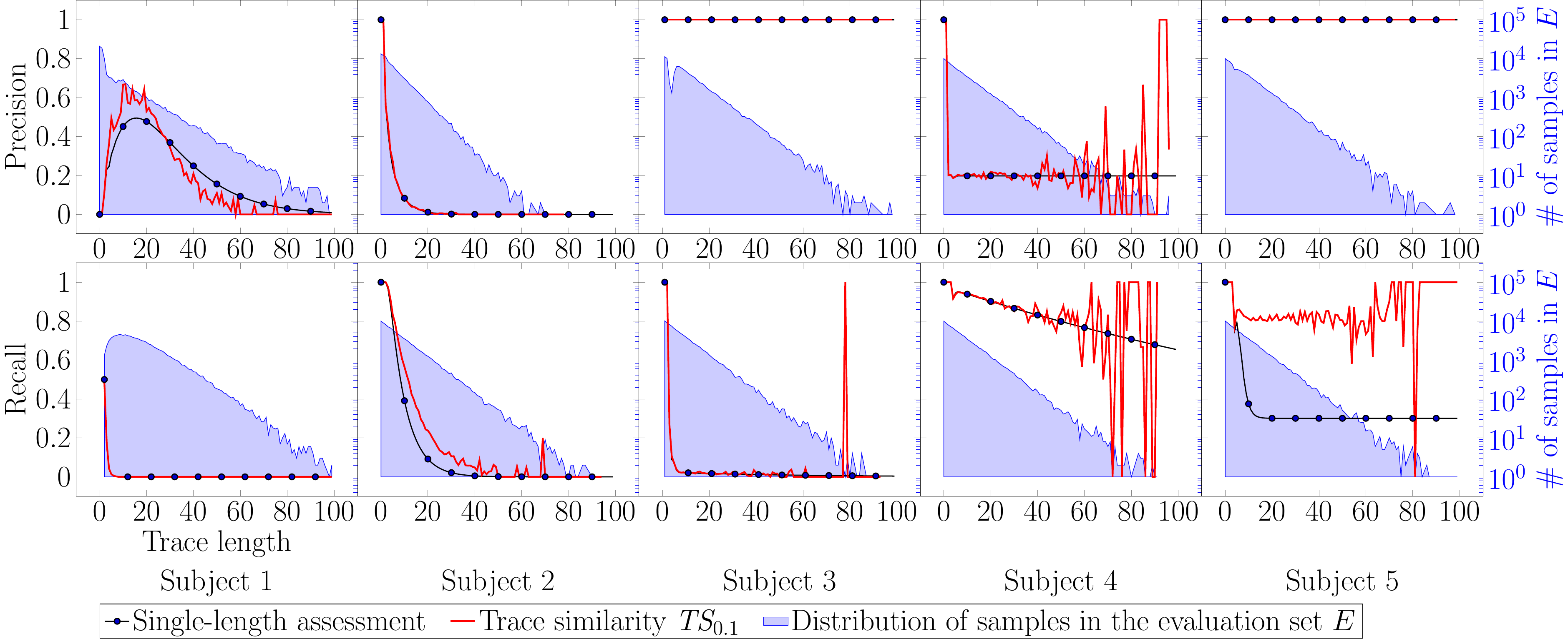}
		\caption{Trace similarity using a random walk with $p_a = 0.1$}\label{fig:comparison-partitioned-p1}
	\end{subfigure}
	\hfill
	\begin{subfigure}[b]{\textwidth}
		\centering
		\includegraphics[width=0.9\textwidth]{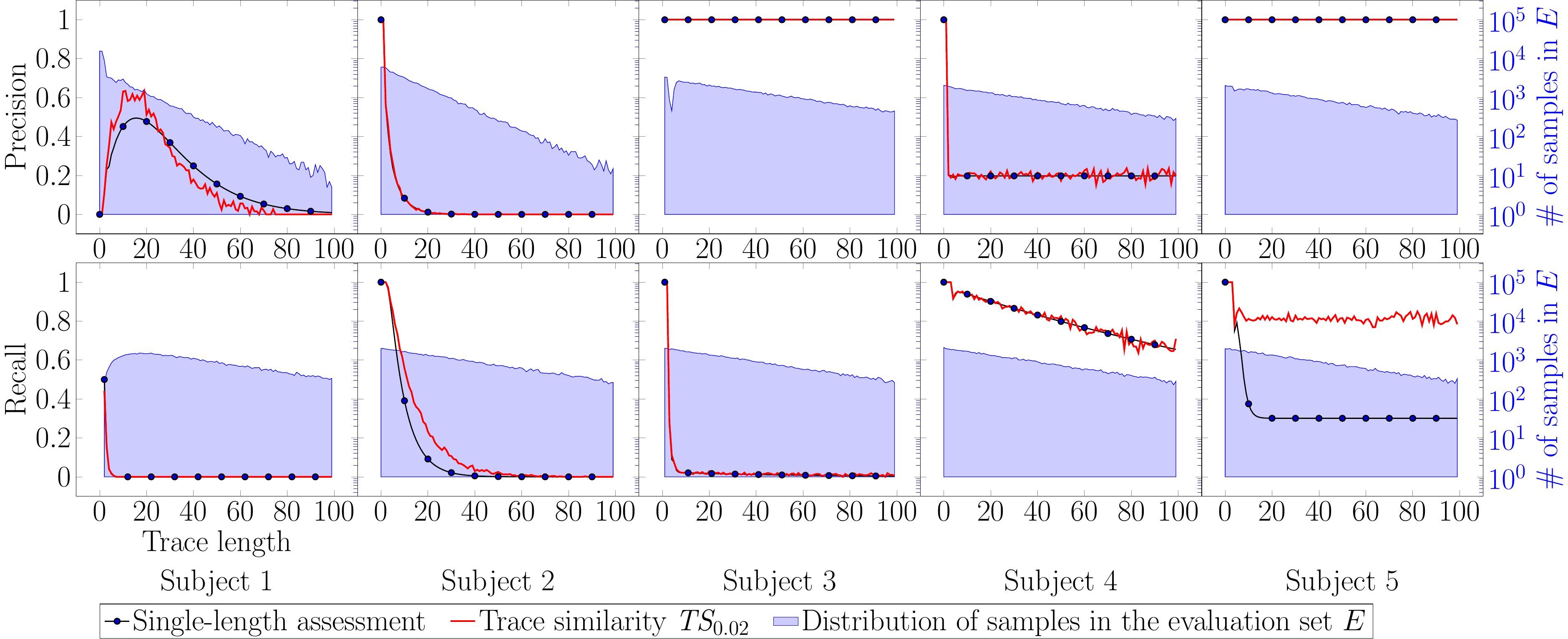}
		\caption{Trace similarity using a random walk with $p_a = 0.02$}\label{fig:comparison-partitioned-p02}
	\end{subfigure}
	   \caption{Comparison of the single-length assessment and trace similarity
       conditioned on the trace length, with three different random walk
       parameters}\label{fig:comparison-partitioned}
\end{figure}

The results for the five representative test subjects previously used in RQ1
are shown in Figure~\ref{fig:comparison-partitioned}.

In each plot, the red line and the black line with markers represent how the
accuracy metric values (left y-axis) computed by trace similarity conditioned on
the trace length and the single-length assessment, respectively, vary depending
on the trace length (x-axis). In addition, the solid area filled in blue
indicates the number of samples (right y-axis) in the evaluation set $E$ for
each trace length (x-axis). Note that, in some of the plots, the red line and
the blue area do not exist after a certain trace length, meaning that the random
walk generated an empty $E$ and no precision and recall values were computed by
trace similarity. 

Based on the results shown in
Figure~\ref{fig:comparison-partitioned}, we can make the following observations.

\begin{itemize}
    \item The distribution of samples highlights how shorter traces are more
    likely to be sampled, and how the distribution of samples depends both on
    the random walk parameters and on the model upon which the random walk is
    performed. For example with $p_a=0.5$
    (Figure~\ref{fig:comparison-partitioned-p5}) the distribution of samples for
    subject S1 contains longer traces, compared to the other test subjects.
    Conversely, with $p_a=0.02$ (Figure~\ref{fig:comparison-partitioned-p02})
    the distribution of samples for the precision assessment of subject S1
    contains shorter traces, compared to the other test subjects.

    \item The random walk can introduce a sampling bias between traces having
    the same length. This is particularly visible in
    Figure~\ref{fig:comparison-partitioned-p02}, in the recall plots of subject
    S5, which show a difference between trace similarity conditioned on the
    length (red line, which converges to 0.8) and the single-length evaluation
    (black line, which converges to 0.3). This observation explains why in
    Figure~\ref{fig:comparison-cumulative}, in the case of subject S5, the
    recall values obtained using standard trace similarity consistently exceed
    the recall values obtained using our cumulative-length assessment, despite
    the fact that the inferred model consistently shows a 0.3 recall value for
    traces having a length greater than 20.
    Note that in the other test subjects, the plots of trace similarity
    conditioned on the trace length, and those of the single-length
    assessment show the same trend, highlighting how this effect is
    model-dependent.
\end{itemize}

One might think that using an assessment method like trace similarity
adjusted to assess the accuracy for long traces could be used to evaluate the
asymptotic behaviour of the model accuracy; however, this approach could give
misleading results. If the asymptotic behavior of the model accuracy converges
to a value that is neither zero nor one, we have observed (in Subject S5) that
trace similarity may converge to a wrong value. If the asymptotic behavior
converges to zero or one (the most common case), it is yet to be proved that the
trace similarity bias would not affect this behaviour. Moreover, although most
models have precision or recall converging to 0 or 1, the speed at which they do
so is relevant for practical purposes (a model accuracy could be asymptotically
zero, but still acceptable for short traces), and this information would be lost
using this approach.

\subsubsection*{Answer to RQ2}
The sampling bias affects the trace similarity result in a way that is hard to
predict a priori, because it depends on both the topology of the model and the
parameters of the random walk. When trace similarity is compared to the
cumulative-length assessment, the conclusion is that the
choice of the parameter values of both assessment methods affect the result to
an extent that does not allow us to reliably determine the model accuracy, thus
preventing a meaningful comparison among different models in terms of accuracy,
or among the accuracy values of the same model measured with different methods.

On the other hand, the single-length assessment is not affected by sampling bias
because it measures the model accuracy for each trace length in the given range,
considering (for each length) all the possible traces. Leveraging this
characteristic, we compared the single-length assessment with trace similarity
conditioned on trace length and analyzed the effect of non-uniform sampling
among traces of the same length. We found that trace similarity can indeed
converge to different and possibly misleading accuracy values due to this
model-dependent bias.
We also examined the distribution of trace length generated by the various
random walk in trace similarity, noticing how shorter traces are 
more likely to be sampled, and how this effect depends on the model topology
and the parameters of the random walk.

\subsection{RQ3: Single-length assessment and model-independent sampling of
$\Sigma^*$} 
The goal of RQ3 is to compare, in terms of precision and recall, the
single-length assessment we proposed and a model-independent sampling of
$\Sigma^*$ inspired by the observation made in
Sections~\ref{sec:sampling-sigma-star} and~\ref{sec:our_approach}. It is possible to
statistically assess precision and recall using a sampling method that generates
traces by combining symbols randomly chosen from the alphabet, rather than
through random walks on models. In practice, however, using this method is
feasible only for short trace lengths, because to evaluate precision
(respectively, recall) it is necessary to generate traces that are accepted by
the inferred (respectively, reference) model. This generation process becomes
infeasible with real-world models and long traces, due to the gap between the
exponential growth of $|\Sigma^n|$ and the slower growth of the size of the
accepted language of length $n$, as the length $n$ is increased. Nonetheless,
this approach has the advantage of controlling the sampling bias (which becomes
model independent), enabling sampling with uniform distribution over a finite
subset of $\Sigma^*$.

Answering
this research question will let us check experimentally whether a statistical
estimation of the model accuracy (in which all the traces having the same length
have the same probability of being sampled) generates the same results as our
single-length assessment. If the results show consensus between the two
assessment methods, they will confirm that our method is a usable alternative to
the statistical evaluation of model accuracy using model-independent sampling of
$\Sigma^*$.

\paragraph{Methodology} 
For each test subject, we computed precision and recall for each trace length
starting at 0 and increasing it until a one-hour timeout was reached.
This was done using Algorithm~\ref{alg:statistical_evaluation_sigma_star}, which
takes as input the reference model $\mathcal{R}$, the inferred model $\mathcal{H}$,
the accuracy metric $m$ to be computed (precision or recall) and 
the trace length $n$ for which to compute it.
The algorithm generates traces without considering the reference or the inferred
model, by concatenating $n$ symbols randomly selected from the alphabet with
uniform distribution. Each generated trace is then tested against the models: if
$m$ is \emph{precision} (respectively, \emph{recall}), the trace is tested
against $\mathcal{H}$ (respectively, $\mathcal{R}$), to determine whether it is
part of the inferred (respectively, reference) language. If this is the case,
the trace is further tested against $\mathcal{R}$ (respectively, $\mathcal{H}$)
to determine whether it is a true positive. This is repeated until 1000
``useful'' traces are generated, i.e., traces that are accepted by $\mathcal{H}$
(respectively $\mathcal{R}$) and therefore contribute to computing the metric.
The target number of useful traces is a compromise between the scalability and the noise of the
evaluation. It was chosen empirically, and deemed appropriate because it gives a
99\% chance that the real precision/recall value is within ±4.08\% of the
measured value. Note that the maximum trace length reached with this model assessment method is
model-dependent, since it depends on the proportion of traces in $\Sigma^*$
accepted by the models under analysis. In fact, it may differ even between
precision and recall on the same test subject, since the proportion may be
different between reference and inferred model.

Finally, we compared the resulting 
precision and recall values with the single-length assessment results.

\begin{algorithm}[t]
	\caption{Statistical assessment of the model accuracy for traces 
    of length $l$, using a model-independent sampling of 
    $\Sigma^l$}\label{alg:statistical_evaluation_sigma_star}
	\SetAlgoLined
	\KwData{Reference model $\mathcal{R}$, inferred model $\mathcal{H}$, 
    alphabet $\Sigma$, trace length $l$, target number of samples $n$, metric $m$ (either \textit{precision} or \textit{recall})}
	\KwResult{Accuracy value}
    \DontPrintSemicolon
    $\mathit{truePositives} \leftarrow 0$\;
    $\mathit{acceptedTraces} \leftarrow 0$\;
    \While{$\mathit{acceptedTraces} < n$}{
        $t \leftarrow \text{randomTrace}(\Sigma, l)$\tcp*[r]{$l$ symbols chosen from $\Sigma$ 
        with uniform distribution}
        \uIf{m = precision $\wedge$ $t \in \mathfrak{L}(\mathcal{H})$}{
            $\mathit{acceptedTraces} \leftarrow \mathit{acceptedTraces} + 1$
        }
        \ElseIf{m = recall $\wedge$ $t \in \mathfrak{L}(\mathcal{R})$}{
            $\mathit{acceptedTraces} \leftarrow \mathit{acceptedTraces} + 1$
        }
        \If{$t \in \mathfrak{L}(\mathcal{R}) \wedge t \in \mathfrak{L}(\mathcal{H})$}{
        	$\mathit{truePositives} \leftarrow \mathit{truePositives} + 1$
        }
    }
    \Return{$\frac{\mathit{truePositives}}{\mathit{acceptedTraces}}$}
\end{algorithm}

\paragraph{Results}

The differences between the model accuracies obtained using the single-length 
assessment and the model-independent sampling of $\Sigma^*$ 
(computed using all the trace lengths for which both results are available)
are summarized in Table~\ref{tab:diff-counting-vs-sampling-sigma-star}.

\begin{table}[tbh] 
    \caption{Mean and standard deviation of the absolute difference in precision and recall 
    evaluated using the single-length 
    assessment and the model-independent sampling of $\Sigma^*$ 
    (values in percentage points).}\label{tab:diff-counting-vs-sampling-sigma-star}
    \begin{tabular}{rcc}
        & $\textit{precision}_{= n}$ & $\textit{recall}_{= n}$ \\ \midrule
    Mean     & \SI{0.44}{\pp} & \SI{0.22}{\pp} \\ 
    $\sigma$ & \SI{0.70}{\pp}  & \SI{0.55}{\pp} \\ \bottomrule
    \end{tabular}
\end{table}

The small differences between the methods' results are caused by the sampling
error intrinsic in any accuracy measure performed using randomly generated
samples, and are within the margin of error expected for the used number of
samples. This confirms that a statistical
estimation in which all the traces of the same length have the same probability
of being sampled generates the same results as our single-length
assessment, in line with your discussion in Section~\ref{sec:our_approach}. 

\begin{figure*}[htbp]
	\centering
    \includegraphics[width=\textwidth]{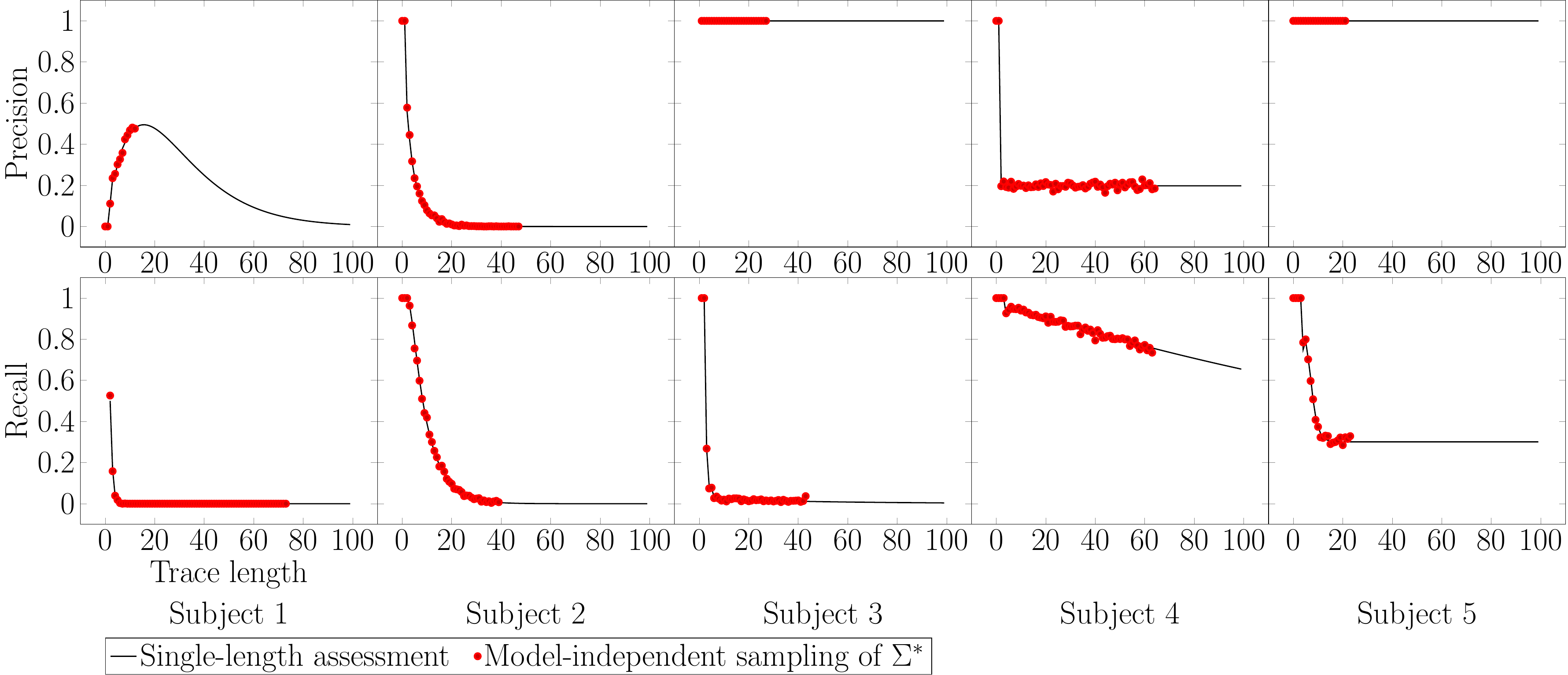}
	\caption{Partitioned comparison of sampling $\Sigma^*$}\label{fig:comparison-sigma-star-partitioned}
\end{figure*}

To have a closer look at the results for individual test subjects,
Figure~\ref{fig:comparison-sigma-star-partitioned} 
shows the results of the same
five representative test subjects previously used in RQ1 and RQ2. The results
obtained using the model-independent sampling of $\Sigma^*$ are represented by
red markers (note the different ranges of trace lengths for which the results
are available, due to the timeout), while the single-length assessment is
represented by the black line.
We remark the two methods exhibit a similar trend also on test subject 5, for
which instead the trace similarity method returned different results due to the
non-uniform probability distribution of traces having the same length, induced
by the random walk.

\subsubsection*{Answer to RQ3}
The precision and recall values obtained 
using the single-length assessment we proposed are consistent with those
obtained using a model-independent sampling of $\Sigma^*$.
This implies that our method is preferable, because it is not limited to short trace lengths, 
and because, being deterministic, is not affected by statistical noise.

\subsection{RQ4: Practical applicability on inferred models} 
This section discusses the practical applicability of our approach by evaluating its
scalability with respect to the complexity of the models under analysis,
measuring the execution time of each assessment on each test subject.

\paragraph{Methodology}

Model complexity is widely studied, and can be measured in a variety of ways,
with different complexity metrics appropriate in different scenarios. We will
evaluate the scalability of our method with respect to two different model
complexity metrics: \emph{deterministic state complexity}~\cite{Yu2001} and
\emph{star height}~\cite{eggan1963transition}.
Specifically, given a regular language $L$, these metrics are defined as
follows. The \emph{deterministic state complexity} is defined as the number of
states of the minimal DFA accepting $L$. We will refer to this measure simply as
the ``number of states'', since all the automata used in this work are in their
minimal form. The number of states is a commonly used model complexity metric in
the field of model inference~\cite{Walkinshaw2016,lo2006smartic}. The \emph{star
height} of $L$ is the minimum star height among all the regular expressions
representing $L$, where the star height of a regular expression is the maximum
nesting depth of Kleene star operators in the expression. Though the start
height is less common, it is a better predictor of the execution time of our
assessment method since this is dominated by the computation of the OGFs.
Referring to the OGF computation algorithm described in
Section~\ref{sec:fastOGF}, it can be seen that an automaton without closed
cycles leads to intermediate OGFs that are easy to compute since they are
polynomials (instead of rational functions) generated using only sums and
products of polynomials. Conversely, all cycles present in the automaton under
analysis must eventually be eliminated during the OGF computation by composing
the intermediate OGFs using sums, products, and quotients of rational functions.
Intuitively, languages with higher star heights are likely to have more closed
cycles in the corresponding DFA representation, increasing the complexity of the
OGF computation, whereas a star height of zero indicates a finite language (thus
with a corresponding DFA with no loops) of which the OGF is a polynomial of
finite degree that is easy to compute.

We should note that the star height is by no means a complete metric; the cost
of each state elimination operation in the OGF computation of
Section~\ref{sec:fastOGF} is hard to predict a priori, because it depends on the
complexity of the rational functions involved. Nonetheless, developing a new
ad-hoc structural complexity metric for the OGF computation is outside the scope
of this work.

Finding the minimum star height of a language is a notoriously hard
problem~\cite{kirsten_2005}. Therefore, in our evaluation, we use an
approximated value obtained by transforming the inferred model in a regular
expression using the Brzozowski and McCluskey
algorithm~\cite{Brzozowski1963fsaToRegex} and then counting the maximum nesting
depth of Kleene star operators in this expression.

Note that both reference and inferred models influence the execution time.
However, in our experiments the size of the inferred 
model is one or two orders of magnitude larger than the size of the reference 
model: therefore we consider the effect of the latter negligible.

Since the differences in execution times between different assessments (single-length and 
cumulative-length) of the same subject are negligible, we give one
execution time \emph{per test subject} rather than \emph{per assessment}.
This is because the execution time is dominated by the computation of the OGFs, 
which is the same for all the assessments on the same test subject. 

\paragraph{Results}

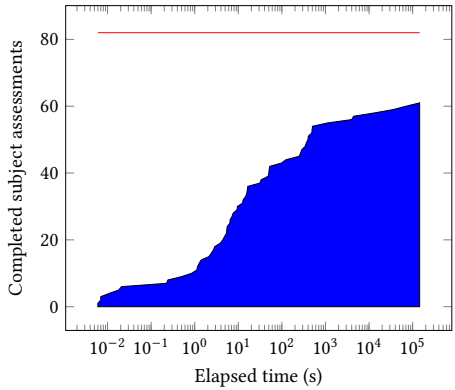
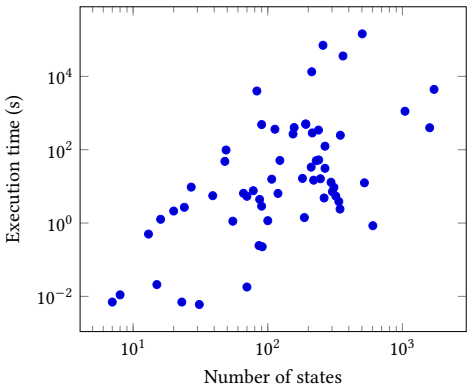
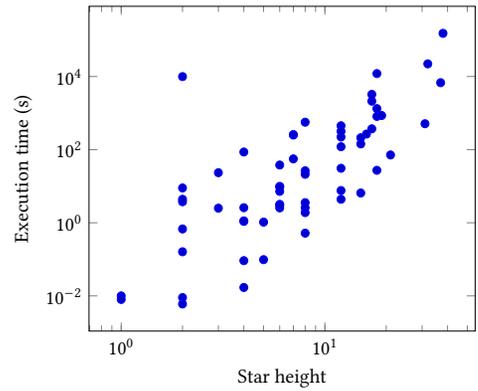
\begin{figure}[tb]
    \centering
    \begin{subfigure}{.447\textwidth}
        \centering
     \begin{tikzpicture}[scale=0.75]
         \begin{semilogxaxis}[compat=1.6,
         xlabel={Elapsed time (s)},
         ylabel={Completed subject assessments}]
             \addplot+[fill=blue,draw=black,no marks]
             table{figures/tikz_generated/data/number-of-evaluated-subjects-vs-time} \closedcycle;
             \addplot[mark=none, red] coordinates {(0.006,82) (144802.691,82)};
         \end{semilogxaxis}
     \end{tikzpicture}
     \caption{Number of subject assessments completed over time.}\label{fig:scalability-num-completed-subjects-vs-time}
    \end{subfigure}\\
    \begin{subfigure}{.447\textwidth}
        \centering
     \begin{tikzpicture}[scale=0.75]
         \begin{loglogaxis}
         [xlabel={Number of states},
         ylabel={Execution time (s)},
         compat=1.6]
             \addplot+[only marks]
             table{figures/tikz_generated/data/evaluation-time-data};
         \end{loglogaxis}
     \end{tikzpicture}
     \caption{Assessment time, depending on the number of states.}\label{fig:scalability-state-compl-vs-time}
 \end{subfigure}\hfill
 \begin{subfigure}{.447\textwidth}
    \centering
 \begin{tikzpicture}[scale=0.75]
     \begin{loglogaxis}
     [xlabel={Star height},
     ylabel={Execution time (s)},
     compat=1.6]
         \addplot+[only marks]
         table{figures/tikz_generated/data/evaluation-time-starh};
     \end{loglogaxis}
 \end{tikzpicture}
 \caption{Assessment time, depending on the star height of the inferred model.}\label{fig:scalability-star-height-vs-time}
\end{subfigure}
    \caption{Scalability of our method}\label{fig:scalability}
\end{figure}

\figurename~\ref{fig:scalability-num-completed-subjects-vs-time} shows the
number of subject assessments completed over time. On one hand, 21 test subjects
assessments did not finish within the timeout, and when the timeout occurred all
of them were in the OGF computation phase. This confirms that the execution time
of our method is dominated by the computation of the OGFs. Among the subject
assessments that did not finish within the timeout, the minimum state count was
68 and the maximum was 2061, while the minimum approximated star height was 13
and the maximum was 304. 
Even though our method takes longer than the trace similarity
  method (which finished all executions within the timeout, with an average of only 168 seconds), 
recall that the results obtained through the trace similarity method are biased, 
as discussed in the previous research questions.
On the other hand, 41 of 82 subjects had an assessment
execution time of less than one minute, showing the practical suitability of the
method. Additionally, a benefit of our method is that the OGFs, once generated,
can be reused to perform further analyses with different parameters with
negligible overhead.

Note that, although the OGF computation is time-consuming in our method,
it is significantly more practical than a naive method 
that enumerates all the possible traces up to a certain length. 
Let us consider a reference model with an alphabet size of 21, 
which is the average alphabet size of the reference models used in our experiments. 
With this alphabet size, assessing the accuracy for traces
up to a modest length of 15 would require evaluating $21^{15}$ traces against
the reference and inferred models. Even under the overly optimistic assumption
that a modern CPU at 4 GHz could test one execution trace against one model
in one clock cycle, this assessment would require more than 1000 years.

\figurename~\ref{fig:scalability-state-compl-vs-time} shows the correlation between 
number of states of the inferred model, and execution time. The largest model 
successfully assessed had 1465 states, and was assessed in \SI{9935}{\s}. 
The Pearson correlation coefficient between execution time and 
number of states was 0.07, highlighting how inadequate is this complexity 
metric to capture the computational complexity of the assessment.

\figurename~\ref{fig:scalability-star-height-vs-time} shows the correlation between 
star height and execution time. The most complex inferred model successfully 
assessed had an approximated star height of 38 (i.e., the maximum nesting depth 
of the Kleene star operator in 
a regular expression accepting the inferred language
was 38), and was assessed in \SI{151000}{\s} --- the longest execution time 
encountered among the test subjects that finished within the timeout.
The Pearson correlation coefficient between star height and execution time was 0.51,
implying that the start height can be a rough indicator of the execution time of our method.

Although the experiments were executed on a multicore system, our current 
implementation is still sequential, thus the performance metrics in this 
section are to be considered for a single core.

\subsubsection*{Answer to RQ4}
Despite the inadequacy of widely adopted model complexity metrics, we report that
our approach was able to assess inferred models representing aspects of actual software
systems, using commodity hardware and despite being at a prototype stage.

\subsection{Threats to validity}\label{sec:threats-to-validity}
There are a number of factors that could impact the validity of our experimental
results.

\emph{Choice of metrics.} 
The choice of the accuracy metrics used in 
this evaluation could threaten the validity of our results if the metrics fail 
to capture adequately the model accuracy.
To mitigate this threat we used precision and recall~\cite{tharwat2020classification}, 
which are well known metrics and have often been used in the area of model 
inference~\cite{Lo2006,Pradel2010,Walkinshaw2016,Krka2014,Polyvyanyy2020,Walkinshaw2008}. 
Moreover, this risk is also partially mitigated by the fact that the proposed assessment method can be adapted 
to compute additional metrics (e.g., specificity or F-measure) without affecting the scalability.

The choice of the metric used to capture the model complexity in the evaluation 
of our method scalability could also pose a construct validity threat. To 
minimize this risk, we provided results using two different complexity metrics: 
model state count and language star height. 

\emph{Reference models used.} 
Using certain reference models could limit the generalizability of our results.
To mitigate this issue, we considered a variety of reference models (41 in total), 
obtained from two different publicly available sources we found in 
the model inference literature~\cite{Pradel2010,Krka2014}.

\emph{Training set generation and model inference setup.}
Due to the lack of a suitable and widely 
accepted model inference benchmark suite containing 
traces of real software executions, we used random 
walks on the reference models to generate the traces 
fed to a model inference method. This could introduce a bias in the training 
set, which may become a bias in the corresponding inferred models.
Similarly, also the choice of the inference method (and its parameter values) 
could introduce a bias in the inferred model.
Nevertheless, the model assessment method proposed in this paper and considered in our evaluation
is independent of the method used for generating inferred models.
Furthermore, we used the same evaluation subjects (i.e., pairs of reference and inferred models) 
for the model assessment methods used in our comparative evaluation, 
yielding fair comparison results. 
Moreover, testing inference methods using traces generated from random walks is common 
practice~\cite{Busany2019,Lo2012,Walkinshaw2013,Walkinshaw2013stamina}.

\emph{Input parameters of other assessment methods.} 
The comparison discussed in RQ2 is affected by the parameters of the considered assessment 
methods, e.g., the characteristics of the random walk used for trace 
similarity. To obtain results that are representative of how the assessment 
methods are is used in practice, our choice of parameter values was guided by 
what is frequently used in the literature. 
Nonetheless, different method parameters would in general lead to different results.

\emph{Statistical noise}. 
Although single-length, cumulative-length, and MBT-based assessment methods are deterministic,
trace similarity involves randomness, which could affect the evaluation results. 
However, due to the low computational cost for generating traces through random walks and checking them against a model,
we were able to perform the trace similarity assessment using at least 
 \num{100000}  samples, ensuring low statistical noise.

\emph{Errors in the implementation.}
We used a prototype implementation of our method, which may contain faults.
Moreover, in order to develop a self-contained solution that does not rely on
external algebra systems such as Wolfram Mathematica~\cite{Mathematica} or
Maple~\cite{Maple}, the algebraic system used to perform operations on rational
functions was implemented by us from scratch: it may also contain faults. To
mitigate the issue, we performed systematic sanity checks by comparing the
language cardinalities obtained using different counting methods.

 \section{Related Work}\label{sec:related-work}

Our method is related to the work done in the area of \emph{model assessment}
and \emph{trace (model) counting}\footnote{This area of research is generally
  called \emph{model counting}. In our context, this name could be confusing,
  since we have already used the term \emph{model} to indicate a representation of
  an aspect of a system. To avoid any confusion, hereafter we refer to it as
  \emph{trace (model) counting}.}. 

\subsection{Model Assessment}

As discussed in Section~\ref{sec:evaluation_with_trace_similarity},
\citet{Lo2006} proposed a method called \emph{trace similarity} for empirically
assessing the accuracy of inferred models against reference models, both in the
form of probabilistic or non-probabilistic finite-state automata. The method
measures the accuracy in terms of \emph{precision} and \emph{recall} using
traces randomly generated from the models. Specifically, for a reference model
$R$ and an inferred model $H$, they defined \emph{recall} as the percentage of
traces generated by $R$ that are accepted by $H$ and \emph{precision} as the
percentage of traces generated by $H$ that are accepted by $R$. They also
proposed an algorithm called \texttt{TraceGen} that generates a set of random
traces from a model with the aim of covering every transition in the model at
least $n$ times, where $n$ is a coverage parameter. \citet{Lo2012} later
extended the trace similarity method to additionally assess the accuracy of
inferred models in terms of \emph{specificity}, which is a metric indicating the
ability to correctly reject illegal behaviors. One common issue of these
methods, as discussed in Section~\ref{sec:evaluation_with_trace_similarity}, is
that the accuracy of inferred models depends on the traces drawn according to a
certain probability distribution imposed by the random trace generation.
Our method resolves this issue by computing precision and recall values that 
consider all the traces, up to a user-defined arbitrarily large maximum length.

As discussed in Section~\ref{sec:evaluation_with_model_based_testing}, the 
aforementioned issue was also acknowledged by \citet{Walkinshaw2008}, which addressed 
it by adapting the W-Method~\cite{Chow1978} (originally developed for model-based 
software testing to generate a set of traces covering all distinguishable runs 
of the model under test) to generate a ``representative'' trace set that covers all 
the model behaviors 
without privileging any specific one. 
Although this solution guarantees to cover all behaviors, it could still return 
misleading results (e.g., an inferred model having fewer counterexamples could 
obtain a lower accuracy value) since it does not consider the number of 
accepted traces affected by each behavior, which is something that our method 
does --- up to a finite maximum trace length set by the user.
Furthermore, as shown in our evaluation results, the MBT-based method is not scalable due to 
the exponential growth of the cardinality of the representative set with 
the number of different states between reference and inferred models. 

Recently, \citet{Polyvyanyy2020} have proposed a framework that compares
reference and inferred models in terms of their languages (i.e., the languages
accepted by the models). While they defined precision and recall similarly to
ours (i.e., Equation~\ref{eq:precision-recall-with-cardinality-cumulative}),
they overlooked the usage of language cardinality, arguing that it is useful
only for finite languages. Instead, they suggested framework instantiations
using \emph{topological entropy} as a language measure which, intuitively,
measures the increase in the variability of the words of the language as their
length goes to infinity. It is not clear
whether, using this measure, an inferred model with fewer counterexamples would
always obtain a higher accuracy value, even when the counterexample languages of
the two inferred models are not in a subset-of relationship. In our method, all
the traces up to the maximum trace length specified by the user are considered,
and they all have the same weight on the result. Moreover, the maximum trace
length parameter can be set either to a value that is relevant for the
application under analysis, or to a very large value, resulting in effectively
computing the asymptotic accuracy.

In the area of formal languages, several works considered the problem of
determining a notion of \emph{distance} for regular languages. 
\citet{parker17regularDistance} investigated extensions of the Jaccard distance
(a popular distance for finite sets) to infinite regular languages. 
A similar approach is followed by \citet{Cui13similarity}.
Similarly to
\cite{Polyvyanyy2020}, both works use topological entropy to handle these
infinite languages, which raises the same concerns mentioned earlier for \cite{Polyvyanyy2020}. The
authors also discuss using a bounded Jaccard distance considering all the traces
up to a finite length. However, precision and recall are much more prevalently used
in the model inference literature~\cite{Krka2014,Lo2012,Lo2006}, and thus we focus on calculating them
using model counting. Nevertheless, our model counting method can be adapted 
to compute the Jaccard distance with respect to a finite evaluation set.

Other than the \emph{language} perspective, where traces (and thus languages)
generated by models matter in assessing the accuracy of inferred models, model
assessment can also take a \emph{structural} perspective, where the model
structure, i.e., how states and transitions are arranged, is considered.
Structural graph similarity measures have been widely used in various fields,
see for example \cite{Champin03GrSim, bunke2000graph}. Moreover, specialized
methods for evaluating inferred models have also been developed.
\citet{Walkinshaw2013} proposed an algorithm called \texttt{LTSDiff} that
compares the structure of two models, identifying missing or superfluous states
and transitions, and computing precision and recall based on these.
\citet{Pradel2010} noticed that structural differences can have a variable
impact on the size of the language difference. To address the issue, they
proposed an approach that counts the number of transitions that are in common
between reference and inferred models after abstracting the models using a
variant of the $k$-tails algorithm. This allows one to identify states that are
\emph{similar}, even when they are not exactly equivalent, with the parameter
$k$ controlling how much imprecision and incompleteness the metrics accept. 
We remark that
the language perspective and the structural one are complementary: a structural
approach is unable to consider the full extent of the impact (i.e., number of
traces affected) of each missing or erroneous state or transition, while a
language approach cannot provide insights about the structural similarity of
the models.

\subsection{Trace (Model) Counting}
The problem of counting the number of words in a regular language has
been investigated in several works. 
A number of approaches use
the \emph{transfer matrix method}~\cite{Flajolet2009,stanley2011enumerativeCombinatorics}, 
which is an analytic combinatorics approach that relies on linear algebraic operations
on the matrix representing the transition relation of the automaton. For
example, \citet{Aydin2015} used it to perform quantitative information flow and
probabilistic analysis of software,
and \citet{Aydin2018} later extended the work including also parametric
constraints, still using the same trace (model) counting approach. While the transfer
matrix method generates the same OGF as our method, as discussed in
Section~\ref{sec:fastOGF},
our method is more scalable, enabling the assessment of larger models.

\citet{Luu2014} proposed an approach to count the number of strings satisfying
a given set of constraints, using \emph{analytic combinatorics}. Constraints are
individually translated to OGF, and later composed. However, both constraint 
translation and composition are not precise, returning an upper and lower bound 
instead.

\citet{Trinh2017} developed a trace (model) counting approach for a class of string 
constraints based on S3P (an SMT solver for strings constraints)~\cite{s3p2016}.
While S3P works by building a reduction tree, reducing the original formula 
into simpler formulas until a satisfying assignment or a contradiction is found,
trace (model) counting is performed by exhaustively building the entire reduction tree,
in which each node is associated with the OGF representing the count for that
node of the tree, and the counts are propagated bottom-up, from the leaf nodes.

Alternative methods to compute the bounded cardinality of a regular
language without using analytic combinatorics have been also developed.
\citet{Kannan95CountingSamplingRegularLang} and \citet{FinkbeinerTorfah14CountingLtl} 
propose dynamic programming
algorithms that, depending on the size and topology of the automata and the
maximum trace length considered, could compute the bounded cardinality more
efficiently than methods based on analytic combinatorics. 
Indeed, analytic
combinatorics methods require an initial high cost for the computation of the
generating function, after which the cardinality of languages for even large
trace lengths can be computed with negligible costs. On the other hand, dynamic
programming does not require a heavy upfront computation, but the cost for
increased trace lengths grow faster. In addition to this discussion, linear
algebraic methods for the computation of the generating function
\cite{Aydin2015} correspond to the solution of a system of linear equations over
a polynomial ring, which may itself be more efficient than both computing the
generating function via state elimination (as proposed in Section
\ref{sec:fastOGF}) or dynamic programming depending on the topology of the
automaton and the sparsity of its transition relation. These methods can be used
interchangeably to compute the bounded cardinality of a regular language within
our framework. Studying the relative efficiency of these methods is beyond the
scope of this paper. However, we notice how the computation of the generating
function might provide additional insights for future work when extending the
study to the asymptotic behavior of the generating function as defined in
\cite{Flajolet2009}.
It is worth noting that also the related problem
of counting certain types of infinite words in $\omega$-regular languages has
been investigated by \citet{FinkbeinerTorfah17Density} to evaluate the ratio of
words satisfying a specific linear-time property.

A related concept in formal languages is the \emph{density} of a regular
language $L$ \cite{Salomaa1978}, i.e., the $\lim_{n\to\infty}$ of the ratio
between the cardinality of the set of traces of length $n$ in $L$, and
$|\Sigma^n|$. This notion can be generalized using at the denominator the cardinality of the set of
traces of length $n$ of an arbitrary regular language $S$,
instead of all the possible traces of length $n$. This generalization,
discussed by \citet{Bodirsky04},  would allow one to compute the asymptotic
behaviour of the model accuracy; however, a known limitation of this approach is that
the limit might not exist. Furthermore, even when the limit exists,
the asymptotic accuracy does not provide any information on the accuracy
for short trace lengths (which may be even more relevant for certain practical applications).

 \section{Conclusions}\label{sec:conclusions}

In the evaluation of newly developed model inference systems, rigorously
assessing the accuracy of the generated inferred models against ground truth
reference models is crucial. In this paper we have highlighted how commonly used
assessment methods provide results that may not reflect the actual accuracy of
the inferred model under analysis: statistical methods are often affected by a
systematic bias caused by overlooking the impact of the random trace generation,
while deterministic methods that guarantee to discover any error in the model
may not correctly capture the number of traces it affects.

To tackle these shortcomings, we have proposed a method to rigorously measure
the accuracy of an inferred model against a ground truth reference model.

First, our method is \textbf{comprehensive}: it considers all the possible traces up
to an arbitrary finite maximum trace length defined by the user. This is in
contrast with other language-based methods that perform the assessment based on
a much smaller subset of traces.

Second, it is \textbf{deterministic}: our method does not rely on random sampling, and
gives exact results. While current statistical methods may provide a more
flexible time/accuracy trade-off, our approach avoids
the convergence limitations of statistical methods, 

Third, it is \textbf{unbiased}: it considers all the traces up to the maximum length, and
each trace has the same weight on the results. This is an advantage over
assessment methods using a random trace generation process to generate a finite
set of traces on which the assessment result is based: the random sampling
induces a probability distribution over the accepted language introducing a bias
in the result, which is not desirable unless it reflects the domain knowledge
about the application being analyzed.

Fourth, it is \textbf{model-independent}: assessing against the same reference
model different models accepting the same language will generate the same
accuracy measurement. This is in contrast with trace similarity, where the
topology of the model affects the sampling, and thus the result.

We have also highlighted how characterizing model accuracy using only a
pair of precision and recall values may be inadequate because an inferred model
can have a variable level of accuracy depending on the \emph{length of the traces} used
in the assessment. As a result, we have proposed an additional assessment
method, measuring the accuracy separately for each trace length, over
a given range.

We have evaluated our assessment methods experimentally and compared ours with 
the currently popular assessment methods, on
reference models previously used in the model inference literature and inferred
models generated using well known model inference methods. The results highlighted the
shortcomings of current assessment methods that are addressed by our solution.
The results also show that our approach is scalable enough to be used
in practice.

As part of future work, we plan to further evaluate our approach on test
subjects from industry, investigate how the accuracy provided by modern model
inference tools changes with the length of the traces considered, and further
discuss the asymptotic analysis of precision and recall
and its usefulness in assessing the model accuracy.

\begin{acks}                            
  The support of the
  \grantsponsor{HiPEDS}{EPSRC Centre for Doctoral Training in High Performance Embedded and Distributed Systems}{https://wp.doc.ic.ac.uk/hipeds/}
  (HiPEDS, Grant Reference 
  \grantnum{HiPEDS}{EP/L016796/1}) is
  gratefully acknowledged.
\end{acks}

\appendix
\section{Appendix: test subjects categorization and
selection}\label{sec:appendix_test_subj_selection} 

In this section we discuss how we categorized the results for the test subjects
in the experimental evaluation, depending on the characteristics of the trends
of the accuracy metric (precision and recall) values over the range of trace
length, and how we used this categorization to choose the five test subjects that
are given as examples in Section~\ref{sec:experimentalEval}. 

To differentiate the trends, we looked at the value of each accuracy metric for
the shortest and the longest available trace length, with the goal of obtaining
a rough indication whether the metric is increasing, decreasing, or constant. 

The values were aggregated in three categories: greater than 0.999 (below
represented with $\mathit{1}$), smaller than 0.001 (below represented with
$\mathit{0}$), and any other value (below represented with $k$). This is
motivated by the observation that the precision and recall values at the
extremes of the trace length range are generally close to either zero or one, and
when this is not the case it is worth considering the subject separately because
it indicates that in
Equation~\ref{eq:precision-recall-with-cardinality-cumulative} numerator and
denominator have the same order of magnitude, which is a relevant characteristic of
the inferred language.

Based on this categorization, we counted the number of test subjects in each
category, obtaining Table~\ref{tab:test-subjects-trends} (e.g. $\mathit{1}
\rightarrow k$ indicates that the value of the accuracy metric value for short
traces is greater than 0.999 and for long traces is between 0.001 and 0.999).

\begin{table}[ht]
	\caption{Number of test subjects for each metric trend}\label{tab:test-subjects-trends}
\begin{tabular}{@{}rccccccccccc@{}}
\toprule
 & & \multicolumn{9}{c}{\textbf{Recall}} \\
 & & $\mathit{0} \rightarrow \mathit{0}$ & $\mathit{0} \rightarrow k$ & $\mathit{0} \rightarrow \mathit{1}$ 
   & $k \rightarrow \mathit{0}$ & $k \rightarrow k$ & $k \rightarrow \mathit{1}$ 
   & $\mathit{1} \rightarrow \mathit{0}$ & $\mathit{1} \rightarrow k$ & $\mathit{1} \rightarrow \mathit{1}$  \\ \midrule
\multirow{9}{*}{\rotatebox[origin=c]{90}{\textbf{Precision}}} 
& $\mathit{0} \rightarrow \mathit{0}$ & 0 & 0 & 0 & 1 & 0 & 0 & 20 & 2 & 0 \\
& $\mathit{0} \rightarrow k$ & 0 & 0 & 0 & 0 & 0 & 0 & 3 & 0 & 0 \\
& $\mathit{0} \rightarrow \mathit{1}$ & 0 & 0 & 0 & 0 & 0 & 0 & 2 & 0 & 0 \\
& $k \rightarrow \mathit{0}$ & 0 & 0 & 0 & 0 & 0 & 0 & 0 & 0 & 0 \\
& $k \rightarrow k$ & 0 & 0 & 0 & 0 & 0 & 0 & 0 & 0 & 0 \\
& $k \rightarrow \mathit{1}$ & 0 & 0 & 0 & 0 & 0 & 0 & 0 & 0 & 0 \\
& $\mathit{1} \rightarrow \mathit{0}$ & 0 & 0 & 0 & 0 & 0 & 0 & 14 & 2 & 2 \\
& $\mathit{1} \rightarrow k$ & 0 & 0 & 0 & 0 & 0 & 0 & 8 & 1 & 0  \\
& $\mathit{1} \rightarrow \mathit{1}$ & 0 & 0 & 0 & 0 & 0 & 0 & 5 & 1 & 0 \\ \bottomrule 
\end{tabular}\end{table}

Looking at table~\ref{tab:test-subjects-trends}, we observed that all the most
common accuracy trends have a recall value starting at 1 and ending at 0. Within
these we selected four subjects, each one having one of the most common precision
value trend. Specifically:
\begin{itemize}
\item Subject 1 was randomly selected among the 20 subjects having precision starting at
		0 and ending at 0. It is worth noting that in all 20 subjects we
		observed that the precision increases to a maximum value, and then
		decreases.
\item Subject 2 was randomly selected among the 14 subjects having both precision and
		recall starting at 1 and ending at 0.
\item Subject 3 was randomly selected among the five subjects having precision starting at 1
		and ending at 1. It is worth noting that in all 5 subjects we observed
		that the precision is actually constant throughout the range.
\item Subject 4 was randomly selected among the eight subjects having precision starting at 1
		and ending at a value between 0 and 1.
\end{itemize}
These trend categories cover 47 of the 61 available test subjects.

In addition, we selected one more subject (Subject 5) showing a recall value
ending at a value greater than zero. This is motivated by the fact that although
this trend is a rare occurrence, it highlights a relevant characteristic of the
inferred model.

\section{Appendix: description of the selected test subjects}\label{sec:selected_subjects_descr}

In test subject 1 
the inferred model precision value starts at zero, reaches a maximum for 
length 10 (47.5\%), and then converges to zero. 
A possible motivation for this behavior is suggested by the large size
of the inferred model: it is possible that the inference approach could
not find any good generalization of the training data, and as a result
the accuracy of the model follows the distribution of lengths of the
traces contained in the training set. In fact, we verified that the most
frequent trace length in the training data is indeed 10.

In test subject 2 
the precision and recall values converge to zero, indicating that 
the growth of the size of the language of true positives is slower than the growth
of the size of the language of both inferred positives and reference positives.

For test subject 3 
the constant precision equal to 1 is explained by the fact that the language of
false positives is empty (which we also manually verified), indicating that the
inferred language is a (non-empty) subset of the reference language.

In test subject 4 the precision converges to a value greater than zero,
indicating that as the trace length is increased, the cardinalities of numerator
and denominator of
Equation~\ref{eq:precision-recall-with-cardinality-cumulative} have the same
order of growth. The same is true for the recall in test subject 5.

Notably, test subject 5 has also an empty language of false positives, as
suggested by the constant precision value equal to 1. Consequently, the inferred
language is a subset of the reference language, and thanks to the asymptotic
value of the recall we can determine that it covers 30\% of the reference
language --- an exceptionally accurate inference result.

\section{Appendix: tradeoff between model size and accuracy}\label{sec:size_and_accuracy_tradeoff_appendix}

Our method allows engineers to evaluate the tradeoff between the size of the inferred
model, which can often be tuned by changing parameters of the inference method,
and its accuracy across a range of trace lengths. In this section we
investigate this tradeoff for the $k$-tails algorithm.

We recall that the $k$-tails algorithm can be intuitively understood as
generalizing a training set by constructing a prefix tree acceptor (PTA)
accepting all the traces in the training set, and then recursively merging the
states having the same \emph{tails of length k}, i.e., the event sequences of
length $k$ that can occur from a state. Lowering $k$ increases the
number of state merge operations, resulting in a more aggressive generalization
of the training set, a smaller inferred model, and a larger inferred
language. Conversely, increasing $k$ causes fewer state merge operations, thus a
more conservative generalization, a larger inferred model, and a generally
smaller inferred language. Values of $k$ greater than the length of the longest
trace in the training set cannot produce any state merge, thus the training data
is not generalized, the inferred model corresponds to the initial PTA, and the
inferred language contains only the training set.

The parameter $k$ affects model accuracy differently for precision and recall.
The accuracy of the initial PTA, and thus of the model
generated with the highest value of $k$, has a precision value equal to one, and
(assuming that the training set is small compared to the target language) a
recall value close to zero. Decreasing $k$ generalizes the training data by
performing more state merge operations, possibly increasing the recall value
by increasing the size of true positive language through correct generalizations,
and decreasing the precision value through incorrect generalizations.

To investigate the effect of $k$ on the model accuracy for different trace
lengths, we followed the same approach used in the generation of the evaluation
subjects for the experimental evaluation
(Section~\ref{sec:experimentalEval_subjects}), but each training set was
processed using the $k$-tails algorithm for $k = 2, 4, 6, 8$, thus obtaining 41
models inferred using 2-tails, 41 using 4-tails, etc. The accuracy of each
inferred model was then evaluated against the corresponding reference model
using the single-length assessment (Section~\ref{sec:our_approach}). 

\begin{figure*}[hbt]
	\centering
    \includegraphics[width=0.85\textwidth]{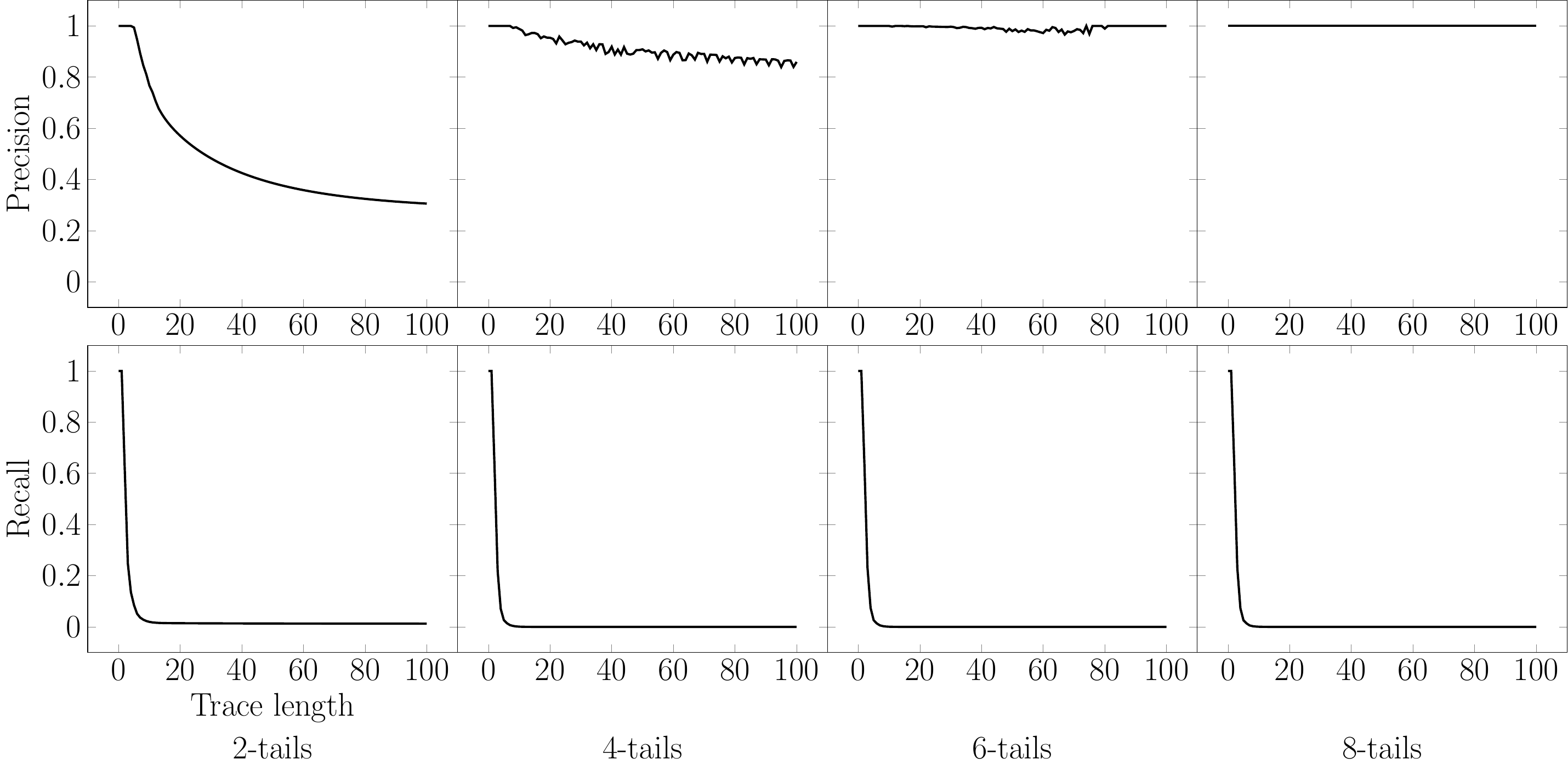}
	\caption{Mean accuracy of models inferred using $k$-tails, with different values of $k$}\label{fig:size_accuracy_tradeoff}
\end{figure*}

The model accuracy results are summarized in
Figure~\ref{fig:size_accuracy_tradeoff}, obtained by computing the mean of all
the precision values (top row) and recall values (bottom row) for the same trace
length ($x$ axis) and the same value of the parameter $k$ (2-tails for the
leftmost column, 8-tails for the rightmost), across all the test subjects. In
accordance with the intuitive summary of $k$-tails presented above, with $k=8$ we
obtained inferred models with the highest precision value (equal to 1 across the
entire trace length range) but low recall value (less than 0.01 for trace
lengths greater than 7). With $k=2$ the precision value is lower (between 1 and
0.28), while the recall value is marginally higher, and stabilizes around 0.012. 
The fluctuations of the precision values visible for 4-tails and 6-tails are
motivated by the fact that for $k \geq 4$ the size of several inferred languages
is relatively small (due to the reduced number of merge operations, as
highlighted before), resulting in widely variable precision values. 
The model sizes are summarized in Table~\ref{tab:size_accuracy_tradeoff}.

This demonstrates that the trade-off between model size and accuracy that can be
achieved by tuning the available parameters of the model inference algorithm
used can be rigorously assessed using our proposed method, and shows that in
the case of $k$-tails it is consistent with the intuitive understanding of how
the algorithm works.

\begin{table}[bth] 
    \caption{Minimum, mean, and maximum state count of models inferred using $k$-tails, with different values of $k$.}\label{tab:size_accuracy_tradeoff}
    \vspace{-9pt}
    \begin{tabular}{cccc}
        \toprule
        		& Min & Avg & Max \\ \midrule
        2-tails & 31 & 258 & 907 \\
        4-tails & 53 & 857.8 & 4405 \\
        6-tails & 58 & 1271.9 & 9637 \\
        8-tails & 60 & 1310.6 & 10506 \\ \bottomrule
    \end{tabular}
\end{table}

\section{Appendix: Correlation between model size and model accuracy.}\label{sec:size_and_accuracy_correlation_appendix}

We have investigated whether the state count of the automata accepting the
language of false positive and false negatives could be useful as a lightweight proxy to estimate the
model accuracy. To do so, we have computed the Pearson correlation coefficient
between the cumulative-length model accuracy at different trace lengths, and the
state count of $\mathcal{A}_{FP}$ and $\mathcal{A}_{FN}$.
Since the inference method affects the characteristics of the inferred model,
this evaluation was performed separately on the models inferred using k-Tails
and MINT.
\begin{table}[bt] 
    \caption{Pearson correlation coefficients between the cumulative-length
    model accuracy at different trace lengths, and the state count of
    $\mathcal{A}_{FP}$ and $\mathcal{A}_{FN}$ (denoted with
    $SC_{\mathcal{A}_{FP}}$ and
    $SC_{\mathcal{A}_{FN}}$).}\label{tab:size_accuracy_correlation}
    \begin{tabular}{r|ccccc|ccccc}
        \toprule
        & \multicolumn{5}{c}{MINT} & \multicolumn{5}{c}{2-tails} \\
        Trace length $n$ & 10    & 20    & 50    & 100   & 200 & 10    & 20    & 50    & 100   & 200   \\ \midrule
        $\textit{precision}_{\leq n}$ vs. $SC_{\mathcal{A}_{FP}}$ & -0.03 & -0.05 & -0.08 & -0.10 & -0.11 & -0.28 & -0.54 & -0.49 & -0.45 & -0.43 \\
        $\textit{precision}_{\leq n}$ vs. $SC_{\mathcal{A}_{FN}}$ &  0.09 &  0.08 &  0.06 &  0.06 &  0.06 & -0.23 & -0.51 & -0.41 & -0.36 & -0.34 \\
        $\textit{recall}_{\leq n}$ vs. $SC_{\mathcal{A}_{FP}}$    & -0.32 & -0.27 & -0.23 & -0.20 & -0.18 & -0.28 & -0.25 & -0.23 & -0.22 & -0.22 \\
        $\textit{recall}_{\leq n}$ vs. $SC_{\mathcal{A}_{FN}}$    & -0.28 & -0.24 & -0.21 & -0.18 & -0.15 & -0.27 & -0.24 & -0.23 & -0.22 & -0.22 \\  \bottomrule
    \end{tabular}
\end{table}

The results are summarized in Table~\ref{tab:size_accuracy_correlation}.
Based on the modest measured correlation and its dependence on the inference
method used for the model generation, we conclude that the state count of the
automata of false positives and false negatives cannot reliably serve as a proxy
of the measured accuracy.

\end{document}